\documentclass[11pt,a4paper,singlecolumn]{article}	

\usepackage[T1]{fontenc}		
\usepackage[utf8]{inputenc}	
\usepackage[english]{babel}	
\usepackage{amsmath}
\usepackage{amsfonts}
\usepackage{amssymb}
\usepackage{graphicx}
\usepackage[font=small,labelfont=bf]{caption}	
\usepackage{etoolbox}	
\patchcmd{\thebibliography}{\section*{\refname}}{}{}{} 

\usepackage{sidecap}

\usepackage[super,comma,sort]{natbib}

\usepackage{lmodern}
 
\usepackage[parfill]{parskip} 	
\usepackage{float}

\usepackage{hyperref}   
\usepackage{amsmath}
\usepackage{amsfonts}
\usepackage{amssymb}
\usepackage{pifont}

\usepackage[table]{xcolor}	

\definecolor{RED}{HTML}{FF0000}
\definecolor{GREEN}{HTML}{00FF00}
\definecolor{BLUE}{HTML}{0000FF}
\definecolor{PINK}{HTML}{FF0DFF} 
\definecolor{TURQUOISE}{HTML}{0BFFFF} 
\definecolor{YELLOW}{HTML}{FFFF02} 
\definecolor{BLACK}{HTML}{000000} 
\definecolor{ORANGE}{HTML}{FF4D01} 
\definecolor{GREY}{HTML}{818181}

\definecolor{green}{HTML}{009900}

\newcommand*{\halffactor}{0.492}	

\usepackage{enumerate}

\usepackage{algorithm}
\usepackage{algorithmic}

\usepackage[table]{xcolor}

\usepackage{amssymb}
\usepackage{rotating}

\usepackage{framed}

\usepackage{setspace}
\author{Robinson Peri\'c}

\begin{document}
\begin{center}
\begin{tiny}
Preprint. Please cite the peer-reviewed journal article, which will be announced at this place when it appears.
\end{tiny}
\hrule
\vspace*{18.0mm}
\begin{LARGE}
\textbf{Analytical Prediction of Reflection Coefficients for Wave Absorbing Layers in Finite-Volume-Based Flow Simulations}
\end{LARGE}
\vspace{12.0mm}

Robinson Peri\'c\footnote{Corresponding author. E-mail-address: robinson.peric@tuhh.de}
\vspace*{1mm}

\begin{scriptsize}Institute for Fluid Dynamics and Ship Theory (M-8), Hamburg University of Technology (TUHH), D-21073 Hamburg, Germany\\ \end{scriptsize}
\end{center}

\vspace*{14mm}
\begin{footnotesize}

{\fontfamily{ptm}\selectfont  ABSTRACT

\vspace*{2mm}
In finite-volume-based flow simulations, absorbing layers are widely used to reduce pressure wave reflections at boundaries of the computational domain. A disadvantage of absorbing layers is that they contain case-dependent parameters; thus the question is how to optimally tune these parameters, so that a desired reduction of reflections can be obtained? As a step towards the answer of these questions, this article presents a theory which predicts reflection coefficients for absorbing layers. The theory is given for 1D-wave propagation and is then extended to 2D and 3D to cover waves of oblique incidence. The theory is validated via flow simulations of regular and irregular pressure waves in air and water, based on Navier-Stokes-type equations and the finite-volume method. Theory predictions and simulation results show good agreement.  
It is demonstrated how the theory can be used to optimally tune the absorbing layer parameters to minimize undesired wave reflection. 
Thus the theory has benefits for a wide range of applications of finite-volume-based flow simulations in industrial practice.

}
\end{footnotesize}
\vspace*{2mm}
\begin{scriptsize}\textbf{Keywords}: Wave absorbing layer, reflection coefficient,  finite-volume method, non-linear flow \end{scriptsize}

\vspace*{12mm}

\section{Introduction}
\label{SECintro}
Computer simulations of pressure wave propagation phenomena are usually  performed on a finite domain, selected as small as possible to minimize the computational effort. Thus it is important to reduce wave reflections which occur  at the domain boundaries, or else these undesired reflections travel back into the solution domain and can lead to substantial errors in the results.

This is especially problematic in finite-volume-based simulations of complex flows, in which pressure wave propagation occurs alogside viscous effects like turbulence and other non-linear phenomena. Such flows are e.g. of interest in fundamental research on acoustic noise generation, and can be computed using flow solvers based on Navier-Stokes-type equations, such as direct numerical simulations of the viscous Navier-Stokes equations (DNS) or large eddy simulations (LES)\cite{REFwang2006,REFcoloniusreview2004}. The present article is intended for such finite-volume-based flow solvers, although application to other solvers is not excluded.

For less complex flows, the governing equations can be simplified and solutions can be obtained more efficiently, e.g. via boundary-element-based potential flow solvers. Then, it is possible to formulate elegant and highly efficient boundary treatments to minimize reflections, such as artificial boundary conditions (also called non-reflecting boundary conditions (NRBC), radiation boundary condition, etc.)\cite{REFcolonius2004,REFmarburg2008,REFgivoli2004,REFisraeli1981}, the supergrid approach\cite{REFcolonius2004,REFcolonius2002}, or perfectly matched layers \cite{REFberenger1994,REFberenger1996,REFmarburg2008,REFcolonius2004,REFhu2008}. 

However, for complex flows as can occur in finite-volume-based flow solvers, many of these sophisticated boundary conditions either cannot be derived, or can produce significant reflection in a manner that is difficult to predict before running the simulation\cite{REFcoloniusreview2004,REFcolonius2004}, or else may not be possible to implement in solvers that do not provide full access to the source code. 
Although the majority of todays (hydro-)acoustic flow problems can be simulated more efficiently using boundary element codes or acoustic analogies, research in finite-volume-based flow solvers has its justification for selected applications, and also for fundamental research\cite{REFmanning2000,REFven2009,REFwang2006,REFcoloniusreview2004}.

Thus how to efficiently and reliably minimize reflections is still an important and urgent question for complex flow simulations. 
As a step towards the answer to this question, this work investigates absorbing-layer-type approaches. 

Absorbing layers (also called forcing layers, relaxation zones, damping zones, sponge layers, porous media layers, buffer regions, etc.) describe a part of the computational domain, usually attached to the domain boundaries, in which source terms are applied to the governing equations. The source terms \textit{gradually force the solution towards  a prescribed  solution}. 
While forcing towards a  steady far-field solution often simply corresponds to a damping of the waves which enter the layer\cite{REFisraeli1981}, forcing towards an unsteady far-field solution can be used to generate waves which propagate from the absorbing layer into the domain, while simultaneously damping waves which enter the absorbing layer\cite{REFkim2012}. 

Absorbing layers can  minimize reflections of several flow phenomena simultaneously when correctly set up, and thus have their benefit for complex flow simulations. Apart from acoustics, absorbing layers are applied e.g. to reduce reflections of free surface waves, K\'{a}rm\'{a}n vortex streets and fully turbulent flows at domain boundaries\cite{REFpark1999,REFchoi2009,REFperic2016,REFwanderley2005,REFandersson2004,REFmanning2000}. Although widely used accross various disciplines, absorbing layers are currently not well understood. This may be partly because they have previously been considered 'ad hoc', suggesting that their reflection behavior is unpredictable\cite{REFcolonius2004,REFmani2012,REFcoloniusreview2004}. In contrast, this work shows that it is indeed possible to analytically predict the reflection from absorbing layers.

The reflection behavior of absorbing layers depends on the forcing strength $\gamma$, which regulates the overall magnitude of the source term, the choice of blending function $b(x)$, which describes how the source term varies within the layer, and the layer thickness $x_{\mathrm{d}}$, which is the smallest distance between domain boundary and entrance to the layer. These parameters are case-dependent, and have to be tuned to provide satisfactory reduction of undesired reflections. 

Generally, benefits of absorbing-layer-type approaches are their often simple implementation and formulation, as well as their flexibility. Even  for  highly nonlinear flows, an absorbing layer with a simple linear damping may successfully reduce undesired reflections. A drawback is the increase in domain size and computational effort.

However, the main problem is that it is not known how to optimally tune the case-dependent parameters to achieve reliable wave absorption\cite{REFberenger1996,REFperic2016,REFmani2012,REFcoloniusreview2004}. It was argued that these "parameters and profiles [...] can only be determined by trail and error"\cite{REFcolonius2004}$ ^{\mathrm{,p.337}} $.
For engineering practice, an approach is needed which can predict the optimum parameter settings with negligible effort \textit{before running the simulation}.

This paper presents such a theoretical approach, which reliably predicts reflection coefficients and damping behavior for any given absorbing layer formulation that can be written in the form given in Sect. \ref{SECabslayer}. 
The theory is implemented in a computer program, which is made publicly available as free software\footnote{The source code and manual can be downloaded from: \url{https://github.com/wave-absorbing-layers/pressure-wave-absorption}}.
The theory predictions are compared to  results from finite-volume-based flow simulations of regular and irregular acoustic wave propagation in gases and liquids.
In Sects. \ref{SECgoveq} and \ref{SECabslayer}, the governing equations and a general absorbing layer formulation are given, which should be applicable to many implementations. Section \ref{SECcr} discusses the calculation of reflection coefficients. 

The theory is derived for the 1D-case in Sect. \ref{SECtheory} and is compared in Sect. \ref{SECres} to results from flow simulations based on the setup described in Sect. \ref{SECsimsetup}, followed by a discussion of the implications of the theory regarding convergence, the choice of blending functions, and the recommended setup in Sect. \ref{SECDiscusstheory}.  
In Sect. \ref{SECapply1Dto2D}, it is found that for many practical two-dimensional (2D) and three-dimensional (3D) wave problems, 1D-theory suffices to optimally tune the absorbing layer parameters. 
Finally, Sect. \ref{SECtheory2D} extends the 1D-theory to 2D and 3D, to predict reflection coefficients for waves entering the absorbing layer at an arbitrary incidence angle $\theta$, which is validated via 2D-flow simulation results. Practical recommendations for setting up absorbing layers are given.

\section{Governing Equations}
\label{SECgoveq}
The governing equations for the simulations are the equation for mass conservation and the three equations for
momentum conservation: 
\begin{equation}
\int_{V} \frac{\partial \rho}{\partial t}  \ \mathrm{d}V + \int_{S} \rho \textbf{v}  \cdot \textbf{n} \ \mathrm{d}S =   \int_{V} \rho q_{\mathrm{c}} \ \mathrm{d}V \quad ,
\label{EQconti}
\end{equation}
\begin{align}
\frac{\mathrm{d}}{\mathrm{d} t} \int_{V} \rho u_{i} \ \mathrm{d}V 
+ \int_{S} \rho u_{i} \textbf{v}  \cdot \textbf{n} \ \mathrm{d}S =  \nonumber \\ 
\int_{S} (\tau_{ij}\textbf{i}_{j} - p\textbf{i}_{i}) \cdot \textbf{n} \ \mathrm{d}S 
+ \int_{V} \rho \textbf{gi}_{i} \ \mathrm{d}V + \int_{V} \rho q_{i} \ \mathrm{d}V \quad ,
\label{EQnavier_stokes}
\end{align}
with volume $V $ of control volume (CV) bounded by the closed surface $\mathrm{S}$, fluid velocity vector \textbf{v}  with the Cartesian components $u_{i}$,  unit vector \textbf{n} normal to $S$ and pointing outwards, time $t$, pressure $p$, fluid density $\rho$, components $\tau_{ij}$ of the viscous stress tensor,  unit vector \textbf{i}$_{j}$ in direction $ x_{j} $, and $ q_{i} $ comprises the  momentum source terms.
 
Since many acoustic wave phenomena are approximately inviscid, the results in this work apply regardless which formulation for $\tau_{ij}$ is chosen or whether it is neglected altogether; this was verified by running selected simulations first with the standard $k$-$\omega$ turbulence model\cite{REFwilcox}, then as laminar simulation (i.e. without any modeling in Eq. (\ref{EQnavier_stokes})), and as inviscid simulation; as expected, no significant differences in wave absorption were encountered.

The energy equation is
\begin{align}
 \int_{V}\frac{\partial \rho E}{\partial t} \ \mathrm{d}V 
+ \int_{S} \rho H \textbf{v}  \cdot \textbf{n} \ \mathrm{d}S =  
- \int_{S} \dot{q}  \cdot \textbf{n} \ \mathrm{d}S 
\label{EQenergy}
\end{align}
with total energy $E = H - p/\rho$, total enthalpy $H = c_{\mathrm{p}}\tau + \frac{1}{2}|\textbf{v}|^{2}$, heat capacity $c_{\mathrm{p}}$ at constant pressure, temperature $\tau$ and heat flux vector $\dot{q}$.

\section{A General Absorbing Layer Approach}
\label{SECabslayer}
In this work, the following absorbing layer formulation is used as source term in Eq. (\ref{EQnavier_stokes})
\begin{equation}
q_{\rm i} = \gamma  b(x)  \left( u_{i,\mathrm{ref}} - u_{i} \right)\quad ,
\label{EQmomdamp}
\end{equation}
with   reference velocity component $u_{i,\mathrm{ref}} $, velocity component $u_{i}  $, forcing strength $\gamma$ and blending function $b(x)$. Outside the absorbing layer holds $q_{\rm i}=0$. In the cases considered in this work, the reference solution is the medium at rest, so $u_{i,\mathrm{ref}} = 0\, \mathrm{\frac{m}{s}}$.

Most existing absorbing layer approaches can be described either directly by or as a slight modification of Eq. (\ref{EQmomdamp}). Thus the application of the results in this work to other existing absorbing layer formulations is straight forward. The source term is directly proportional to the velocity, since this provides satisfactory damping for a wider range of wave frequencies than source terms proportional to the velocity squared, as discussed in \cite{REFperic2016}.

\begin{figure}[H]
\begin{center}
\includegraphics[width=\halffactor\linewidth]{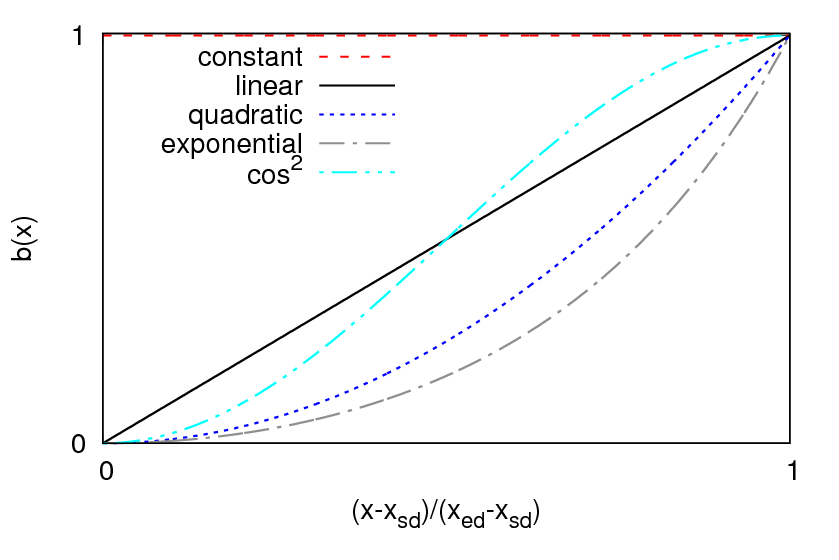}
\end{center}
\caption{Different blending functions $b(x)$ over location in absorbing layer} \label{FIGblend}
\end{figure}

The forcing strength $\gamma$ with unit  $\left[ \frac{1}{\mathrm{s}} \right]$ regulates how strong the solution at a given cell is forced against the reference solution. 
The blending term $b(x)$ regulates the distribution of the source term over the domain, where $x$ is the wave propagation direction.  Many different types of blending functions can be applied. Common choices are constant blending
\begin{equation}
b(x)=1 \quad ,
\label{EQblendCONST}
\end{equation}
linear blending
\begin{equation}
b(x) = \frac{x - x_{\rm sd}}{x_{\rm ed} - x_{\rm sd}} \quad ,
\label{EQblendLIN}
\end{equation}
quadratic blending
\begin{equation}
b(x) = \left(\frac{x - x_{\rm sd}}{x_{\rm ed} - x_{\rm sd}}\right)^{2} \quad ,
\label{EQblendQUAD}
\end{equation}
cosine-square blending
\begin{equation}
b(x) = \cos^{2}\left( \frac{\pi}{2} +\frac{\pi}{2}\frac{x - x_{\rm sd}}{x_{\rm ed} - x_{\rm sd}} \right) \quad ,
\label{EQblendCOS2}
\end{equation}
 or exponential blending such as
\begin{equation}
b(x) = \left( \frac{e^{\left( \frac{x - x_{\rm sd}}{x_{\rm ed} - x_{\rm sd}} \right)^{2}} - 1}{e^{1} - 1} \right) \quad ,
\label{EQblendEXP}
\end{equation}
with start coordinate $ x_{\rm sd} $,  end coordinate $ x_{\rm ed} $, and thickness $x_{\mathrm{d}}=| x_{\rm ed} -x_{\rm sd}|$ of the absorbing layer. These blending functions are illustrated in Fig. \ref{FIGblend} and will be used in the following sections.
 Though so far the optimum blending function is not known, several investigations showed that higher order blending functions are more effective than constant or linear blending. This was attributed to the smoother blending-in of the source term \cite{REFisraeli1981,REFmani2012,REFperic2016,REFkim2014}.
As shown in \cite{REFperic2016}, parameters $\gamma$ and $x_{\mathrm{d}}$ scale with angular wave frequency $\omega$ and wavelength $\lambda$ as 
\begin{equation}
\gamma \propto \omega \quad , \quad x_{\mathrm{d}} \propto \lambda \quad .
\label{EQscaleGammaXd}
\end{equation}
By this scaling the present results can easily be applied to waves of any frequency.

\section{Determining Reflection Coefficient $C_{\mathrm{R}}$}
\label{SECcr}

The reflection coefficient $C_{\mathrm{R}}$ is the amplitude of the reflected wave divided by the amplitude of the original wave. 
For a continuous regular wave train in 1D, $C_{\mathrm{R}}$ can be computed as in \cite{REFursell1960} via maximum $u_{\mathrm{max}}$ and minimum $u_{\mathrm{min}}$ velocity, which occur during the last simulated period for all cells outside the absorbing layer which are within a given distance (here:  $2\lambda$) to the absorbing layer:
\begin{equation}
C_{\mathrm{R}} = \frac{|u_{\mathrm{max}}|-|u_{\mathrm{min}}|}{|u_{\mathrm{max}}|+|u_{\mathrm{min}}|} \quad .
\label{EQcr}
\end{equation}
Alternatively, for cases with short wave trains in 1D, 2D, or 3D one can determine $C_{\mathrm{R}}$ based on the energy in the domain at time $t=t_{\mathrm{end}}$, when the simulation is finished 
\begin{equation}
C_{\mathrm{R}} = \sqrt{\frac{(E_{\mathrm{kin}}+E_{\mathrm{pot}}) |_{\gamma=\gamma_{i},\ t=t_{\mathrm{end}}}}{ (E_{\mathrm{kin}}+E_{\mathrm{pot}})|_{\gamma=0\ t=t_{\mathrm{end}}}}}
\quad ,
\label{EQCrB2}
\end{equation}
where $\gamma_{i}$ is the forcing strength used in the simulation, and the kinetic and potential energies are 
\[E_{\mathrm{kin}} = \int_{V}\frac{1}{2} \rho |\mathbf{u}|^{2}\ \mathrm{d}V  \quad , \quad E_{\mathrm{pot}} = \int_{V}\frac{1}{2} \frac{p^{2}}{\rho c^{2}}\ \mathrm{d}V  \quad ,\]

with density $\rho$, domain volume $V$, velocity vector $\mathbf{u} = (u,w)^{\mathrm{T}}$, pressure $p$ relative to reference pressure, phase velocity $c=\lambda/T$ of the wave, and control volume $V$; see e.g. \cite{REFlerch2009} for a detailed derivation. 
The domain size and simulation duration should be chosen so that at $t=t_{\mathrm{end}}$ the whole wave train has passed through the absorbing layer once, which was fulfilled in all calculations in this work.

\section{1D-Theory}
\label{SECtheory}
The one-dimensional wave equation takes the form
\begin{equation}
\chi_{tt}=  c^{2} \chi_{xx}\quad .
\label{EQwave}
\end{equation}
with particle displacement $\chi$, velocity $u=\chi_{t}$, and speed of sound $c$. It describes waves e.g. in an ideal gas, liquid or solid under isothermal conditions. A detailed derivation can be found in textbooks such as\cite{REFfeynman2013}.

The damping of waves in Eq. (\ref{EQwave}) can be achieved by applying a  'classical' absorbing layer, inside which the wave equation takes the form
\begin{equation}
 \chi_{tt}=  c^{2} \chi_{xx} + q  \quad ,
\label{EQwavewithdamping}
\end{equation}
with in analogy to Eq. (\ref{EQmomdamp}) 
\begin{equation}
q = \gamma  b(x) (\chi_{t,\mathrm{ref}} - \chi_{t})\quad ,
\label{EQmomdampTheory}
\end{equation}
with forcing strength $\gamma$, blending function $b(x)$ and reference velocity $\chi_{t,\mathrm{ref}} $. In this work,  $\chi_{t,\mathrm{ref}} = 0\, \mathrm{\frac{m}{s}}$, which corresponds to Mach number $\mathrm{Ma}=0$.

Assume that, for a wave propagating in positive $x$-direction,  displacement $\chi$ can be written in the complex plane as
\begin{equation}
 \chi = A_{0} \exp {i(-\omega t + kx)} \quad ,
\label{EQchi}
\end{equation}
with particle displacement amplitude $A_{0}$, angular wave frequency $\omega = 2\pi/T$, wave period $T$, wave number $k=2\pi/\lambda$, and wavelength $\lambda$.

Inserting Eq. (\ref{EQchi})  into  Eq. (\ref{EQwave}) gives the wave number \textit{outside the absorbing layer}
\begin{equation}
k= \sqrt{\frac{\omega^{2} }{c^{2}} }  = \frac{\omega}{c} \quad .
\label{EQk}
\end{equation}

For constant blending $b(x)=b$, inserting Eq. (\ref{EQchi}) into  Eq. (\ref{EQwavewithdamping})  gives the wave number \textit{inside the absorbing layer}
\begin{equation}
 k= \sqrt{\frac{\omega^{2}}{c^{2}} + i \frac{  \omega \gamma  b(x) }{c^{2}} } \quad .
\label{EQkstar}
\end{equation}
Thus  inside the absorbing layer, the wave number contains an additional imaginary part which  damps the wave amplitude but does not change the wavelength.

\begin{figure}[H]
\begin{center}
\includegraphics[width=\halffactor\linewidth]{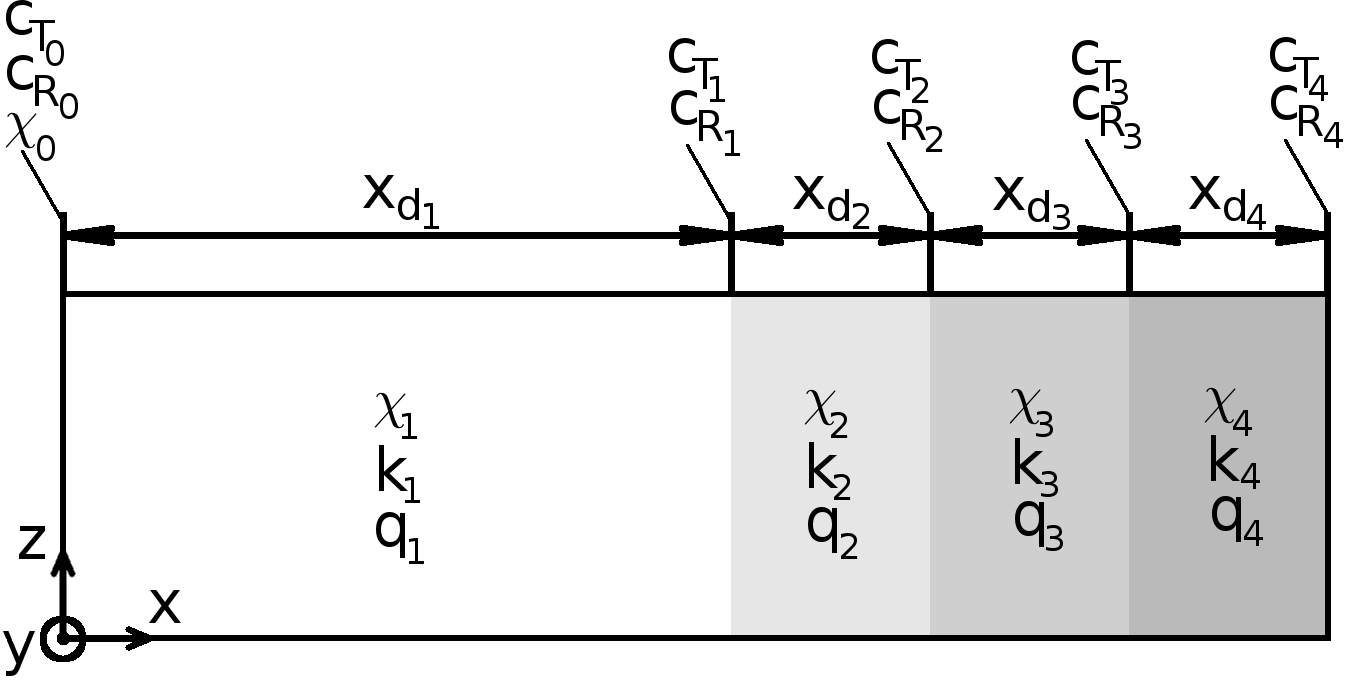}
\end{center}
\caption{Example of a solution domain decomposed into $4$ zones} \label{FIGdomwith4xdzones}
\end{figure}

In this work, the approach for constructing an analytical solution to the problem of determining the reflection coefficient for a wave entering an absorbing layer according to Eqs. (\ref{EQwave}) and (\ref{EQwavewithdamping}) is the following.  The solution domain is discretized into a finite number of zones as illustrated in Fig. \ref{FIGdomwith4xdzones}. Each zone $j$ corresponds to an absorbing layer, within which the particle displacement is given by $\chi_{j}$ and the complex wave number $k_{j}$ has a constant value. For this, the blending function $b(x)$, which can be any function (see e.g. Fig. \ref{FIGblend}),  is evaluated at the zone center, to obtain piece-wise constant blending:
\begin{equation}
 k_{j}= \sqrt{\frac{\omega^{2}  + i \omega \gamma  b( \sum_{n=1}^{j-1} x_{\mathrm{d}_{n}} + \frac{1}{2}x_{\mathrm{d}_{j}}) }{c^{2}} } \quad ,
\label{EQkstar}
\end{equation}
with thickness $x_{\mathrm{d},j}$ of zone $j$; $x_{\mathrm{d},j}$ is equivalent to the size of the zone in $x$-direction. Thus the damping is constant within every zone.
Reflection and transmission may occur at every interface between two zones. 

The benefit of this approach is that even discontinuous damping and the influence of the discretization  can be considered.
With increasing resolution, the theoretical results are expected to converge to the solution of the continuous problem. 
The latter is not derived here, since for practical purposes only the analytical solution to the discretized problem is of interest. In this manner, the problem remains linear and the solution can be derived as follows.

Consider a wave propagating in positive $x$-direction. The wave is generated at $x=0$ following the coordinate system in Fig. \ref{FIGdomwith4xdzones}. Let the particle displacement at $x=0$ be 
\begin{equation}
  \chi_{0} = A_{0} \mathrm{e}^{-i\omega t} \quad ,
\label{EQchi00}
\end{equation}
with a displacement amplitude $A_{0}$, angular wave frequency $\omega=2\pi/T$, wave period $T$ and time $t$. Set the transmission coefficient $C_{\mathrm{T}_{0}} = 1$ and the  reflection coefficient $C_{\mathrm{R}_{0}} = 0$, thus the 'inlet' boundary is perfectly transparent, and waves propagating through it in negative $x$-direction will be fully transmitted without reflection. 

For illustration, a domain with $4$ zones is  depicted in  Fig. \ref{FIGdomwith4xdzones}. Let the wave number  $k_{1}=2\pi/\lambda_{1}$ within zone $1$ equal the wave number $k_{0}=2\pi/\lambda_{0}$ of the  wave generated at $x=0$, where  $\lambda_{0}$ and  $\lambda_{1}$ are the corresponding wavelengths. This means, that within the first zone  $0 \leq x \leq x_{\mathrm{d}_{1}}$, there is no wave damping, i.e. $q(x) = 0$. At the end of the domain, i.e. at $x =\sum_{n=1}^{4}  x_{\mathrm{d}_{n}} $, the boundary is perfectly reflecting (a typical 'wall boundary condition' in computational fluid dynamics), so the transmission coefficient $C_{\mathrm{T}_{4}} = 0$ and the  reflection coefficient $C_{\mathrm{R}_{4}} = 1$.
 Within each zone $j$ in zones $2$ to $4$, the damping is constant (i.e. $q(x)=q_{j}=-\gamma b( \sum_{n=1}^{j-1} x_{\mathrm{d}_{n}} + \frac{1}{2}x_{\mathrm{d}_{j}}) \chi_{t,j}$), but $q_{2}$ to $q_{4}$ may be of different magnitude, which is illustrated through the different shading of the zones.

By requiring that the particle displacements and velocities must be continuous at every interface between two adjacent zones, as they should be at the interfaces between two zones in a flow simulation, the analytical solution is obtained.

For a domain with $j$ zones, the general solution for the particle displacement $\chi_{j}(x)$ within zone $j>0$ can be written as a sum of a right-going (incoming) and a left-going (reflected) wave component
\begin{align}
 \chi_{j} =  \chi_{0}  \left( \prod_{n=0}^{j-1} C_{\mathrm{T}_{n}} \right) \biggl[ \mathrm{e}^{i \left( \sum_{n=1}^{j-1} k_{n} x_{\mathrm{d}_{n}} + k_{j} \left( x -  \sum_{n=1}^{j-1} x_{\mathrm{d}_{n}  }   \right) \right)} \nonumber \\ 
  - C_{\mathrm{R}_{j}} \mathrm{e}^{i \left(  \sum_{n=1}^{j-1} k_{n} x_{\mathrm{d}_{n}}  + k_{j}2x_{\mathrm{d}_{j}} - k_{j} \left( x -  \sum_{n=1}^{j-1} x_{\mathrm{d}_{n} }   \right) \right)} 
\biggr]  \quad ,
\label{EQXjgeneral}
\end{align} 
with $ \chi_{0} $ according to Eq. (\ref{EQchi00}).

\textbf{Requirement 1:} At the interface between zones $j$ and $j+1$, the solution for  the zone on the left, i.e.  $\chi_{j}$, must equal  the  solution for the zone on the right, i.e.  $\chi_{j+1}$: 
\begin{align}
 \left[ \chi_{j} =  \chi_{j+1} \right]_{x=\sum_{n=1}^{j} x_{\mathrm{d}_{n}}} \quad .
 \label{EQBC1}
\end{align} 

Inserting Eq. (\ref{EQXjgeneral}) into Eq. (\ref{EQBC1}), dividing by 
\[   \chi_{0}  \left( \prod_{n=0}^{j-1} C_{\mathrm{T}_{n}} \right)\biggl[ \mathrm{e}^{i \left( \sum_{n=1}^{j-1} k_{n} x_{\mathrm{d}_{n}} \right) }  \biggr] \quad ,\]

and rearranging for $ C_{\mathrm{T}_{j}}$ gives

\begin{framed}
\begin{align}
 C_{\mathrm{T}_{j}} =  \frac{1  - C_{\mathrm{R}_{j}}}{ 1 - C_{\mathrm{R}_{j+1}}  \mathrm{e}^{i \left(  k_{j+1}2x_{\mathrm{d}_{j+1}}  \right) }} 
 \quad .
\label{EQCt}
\end{align} 
\end{framed}

\textbf{Requirement 2:} At the interface between zones $j$ and $j+1$, the  spatial derivative with respect to $x$ of the solution for  the zone on the left, i.e.  $\partial \chi_{j}/\partial x = \chi_{x,j}$, must equal the spatial derivative with respect to $x$ of the solution for the zone on the right, i.e.  $\partial \chi_{j+1}/\partial x = \chi_{x,j+1}$: 
\begin{align}
 \left[ \chi_{x, j} =  \chi_{x, j+1} \right]_{x=\sum_{n=1}^{j} x_{\mathrm{d}_{n}}} \quad .
 \label{EQBC2}
\end{align} 
Inserting the spatial derivative of Eq. (\ref{EQXjgeneral}) into Eq. (\ref{EQBC2}), dividing  by 
\[   i\chi_{0}  \left( \prod_{n=0}^{j-1} C_{\mathrm{T}_{n}} \right)\biggl[ \mathrm{e}^{i \left( \sum_{n=1}^{j-1} k_{n} x_{\mathrm{d}_{n}} \right) }  \biggr] \quad ,\]
inserting $ C_{\mathrm{T}_{j}}  $ according to Eq. (\ref{EQCt}) and introducing
\begin{align}
\beta_{j+1} = \frac{  1 +  C_{\mathrm{R}_{j+1}}  \mathrm{e}^{i  \left( k_{j+1} 2 x_{\mathrm{d}_{j+1}} \right) }  }{  1 -  C_{\mathrm{R}_{j+1}}  \mathrm{e}^{i  \left( k_{j+1} 2 x_{\mathrm{d}_{j+1}} \right) } } 
\label{EQbetajp1}
\end{align} 
gives 
\begin{framed}
\begin{align}
  C_{\mathrm{R}_{j}} = \frac{ k_{j+1}\beta_{j+1} -  k_{j}   }{ k_{j+1}\beta_{j+1} + k_{j} }
 \quad .
\label{EQCRj}
\end{align} 
\end{framed}

For practical purposes, mainly the 'global' reflection coefficient $  C_{\mathrm{R}} $ is of interest, which is the ratio of the amplitude of the wave, which is reflected back into the solution domain, to the amplitude of the wave, which enters the absorbing layer. 
  Reflection coefficient $  C_{\mathrm{R}} $ corresponds to the magnitude of the $  C_{\mathrm{R}_{j}} $ at the interface to the damping layer,  which depends on all $  C_{\mathrm{R}_{j}} $ inside the whole absorbing layer. So if the absorbing layer starts at zone $1$, then
\begin{framed}
 \begin{align}
   C_{\mathrm{R}} = | C_{\mathrm{R}_{1}} | = \sqrt{\mathrm{Re}\{C_{\mathrm{R}_{1}}\}^{2} + \mathrm{Im}\{C_{\mathrm{R}_{1}}\}^{2}}
 \quad ,
\label{EQCRglobal}
\end{align} 
\end{framed}
where $ \mathrm{Re}\{ X\} $ and $ \mathrm{Im}\{ X\} $ denote the real and the imaginary part of the complex number $X$.

\section{Simulation Setup}
\label{SECsimsetup}

In Sect. \ref{SECres}, simulations of one-dimensional (1D) wave-propagation are performed on a  computational domain with length $L_{\mathrm{x}}$. The domain is illustrated in Fig. \ref{FIGdomabslayer}. 
The coordinate system has its origin at $x=0\, \mathrm{m}$. 
At  boundary $x=0\, \mathrm{m}$, a sinusoidal pressure fluctuation is prescribed to produce a continuous regular wave of period $T=2.\overline{27}\cdot 10^{-3}\, \mathrm{s}$. This corresponds to modern standard concert pitch A $  440\, \mathrm{Hz} $.
The wave propagates in positive $x$-direction, is partially reflected, absorbed and transmitted at each cell within the absorbing layer, and the remaining wave is fully reflected at the wall boundary $x=L_{\mathrm{x}}$. 
For convenience, the simulations are run  quasi-one-dimensional, i.e. they consist of a uniform Cartesian three-dimensional grid, with 1 cell in $y$- and $z$-direction and symmetry boundary condition for the $y$- and $z$-normal boundaries; thus all gradients in $y$- and $z$-direction are zero.

In Sects. \ref{SECapply1Dto2D} and \ref{SECres2D}, simulations of two-dimensional (2D) and three-dimensional (3D) wave-propagation are performed on solution domains shown in Figs. \ref{FIGB2P}, \ref{FIGB3P} and \ref{FIGB5p}. The wave generation is discussed in the corresponding sections. The 2D-simulations are run  quasi-two-dimensional, i.e. they consist of a uniform Cartesian three-dimensional grid, with 1 cell in $y$-direction and symmetry boundary condition for the $y$-normal boundaries. 

All simulations are performed using the commercial flow solver STAR-CCM+ (version 8.02.008-R8) by Siemens (formerly CD-adapco) based on the governing equations given in Sects. \ref{SECgoveq}. The absorbing layer formulation is based on Sect. \ref{SECabslayer}. The implicit unsteady segregated solver is used.
All approximations are of second order in time and space, and under-relaxation is $0.8$ for velocities, $0.2$ for pressure and $0.9$ for energy. The initial conditions are $p=0$, $u_{i}=0$ and  $\rho=\rho_{{ref}}$, the reference density of the fluid. 
 Unless stated otherwise, the wave is discretized by $\geq 30$ cells per wavelength and the time step is $\geq T/100$, with $8$ outer  iterations  per time step. Detailed information on finite-volume-based flow simulations can be found e.g. in \cite{REFferzigerperic}.

Simulations are performed for liquid  water using the IAPWS model\cite{REFiapws}  and solving the energy Eq. (\ref{EQenergy}) in a segregated manner for temperature $\tau$, with reference temperature $\tau_{\mathrm{ref}}=300\, \mathrm{K}$. Further, simulations are performed for  an ideal gas with speed of sound $c=291.5\, \mathrm{\frac{m}{s}}$, which corresponds to air at temperature $\approx -62^\circ C$ or to an appropriate mixture of oxygen and carbon dioxide at room temperature. The gas is considered isothermal, so  Eq. (\ref{EQenergy}) is not solved. For simplicity, these fluids will be denoted 'water' and 'ideal gas' in the following.
The simulations are run with different absorbing layer parameters, such as forcing strength $\gamma$, layer thickness $x_{\mathrm{d}}$, or blending function $b(x)$.

\section{1D-Results}
\label{SECres}
\begin{figure}[H]
\begin{center}
\includegraphics[width=\halffactor\linewidth]{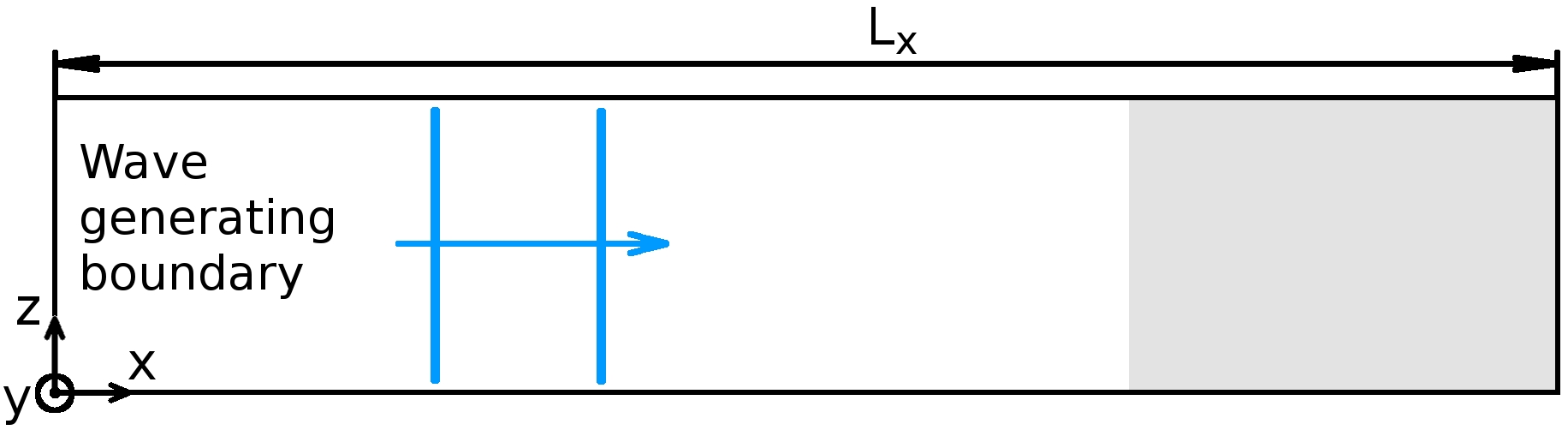}
\end{center}
\caption{Computational domain with wave-maker at $x=0\, \mathrm{m}$ and absorbing layer (shaded gray) attached to domain boundary at $x=L_{\mathrm{x}}=9\lambda$} \label{FIGdomabslayer}
\end{figure}

The theory presented in Sect. \ref{SECtheory} is compared to flow simulation results for 1D regular and irregular pressure waves in an ideal gas and in liquid water. First, results for regular waves are presented. The waves have period $T=2.\overline{27}\cdot 10^{-3}\, \mathrm{s}$ and the amplitude of the pressure fluctuations at the wave-maker is $10\, \mathrm{Pa}$ (ideal gas) or $1000\, \mathrm{Pa}$ (water). The setup is described in Sect. \ref{SECsimsetup}. The computational domain is sketched in Fig. \ref{FIGdomabslayer}. An absorbing layer based on Eq. (\ref{EQmomdamp}) is used in the simulation.

On a single core $2.6\, \mathrm{GHz}$ processor, the time to run a simulation was in the order of $1\, \mathrm{min}$.
The reflection coefficient $C_{\mathrm{R}}$ is computed via Eq. (\ref{EQcr}).
The theory is evaluated for same wavelength, period and absorbing layer parameters in Eq. (\ref{EQmomdampTheory}) as in the simulations. The theory predictions are given for an absorbing layer subdivided into $200$ zones.

\begin{figure}[H]
\begin{center}
\includegraphics[width=\halffactor\linewidth]{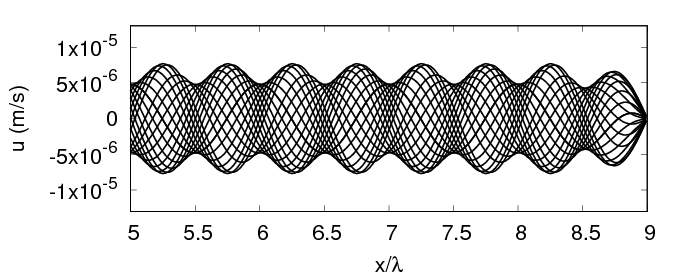}\includegraphics[width=\halffactor\linewidth]{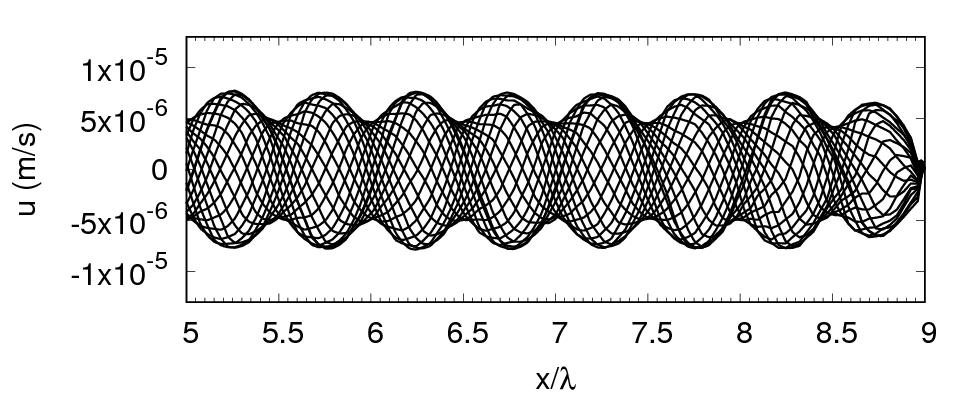}\\
\includegraphics[width=\halffactor\linewidth]{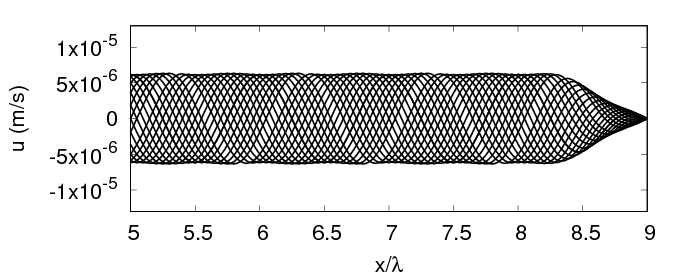}\includegraphics[width=\halffactor\linewidth]{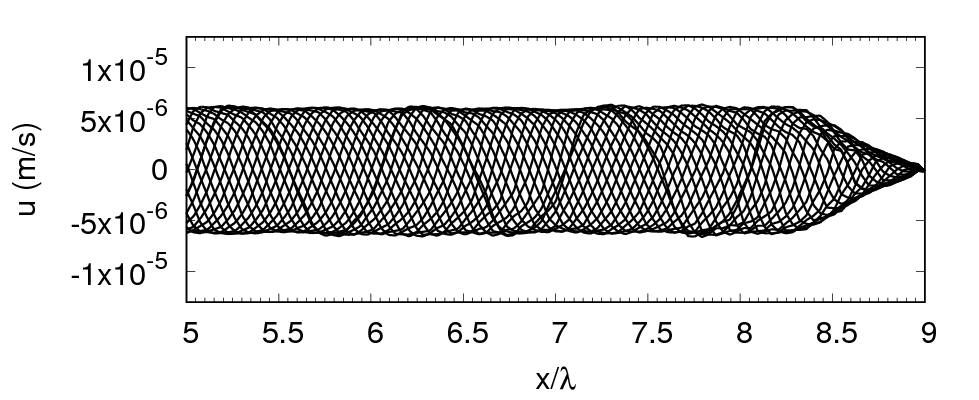}\\
\includegraphics[width=\halffactor\linewidth]{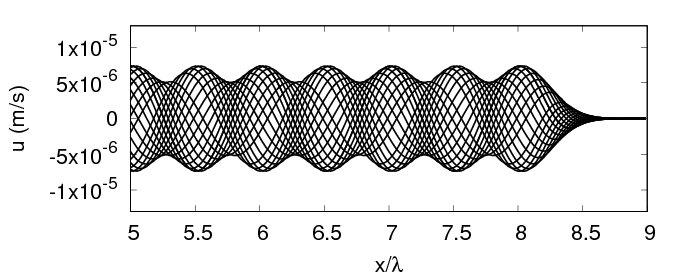}\includegraphics[width=\halffactor\linewidth]{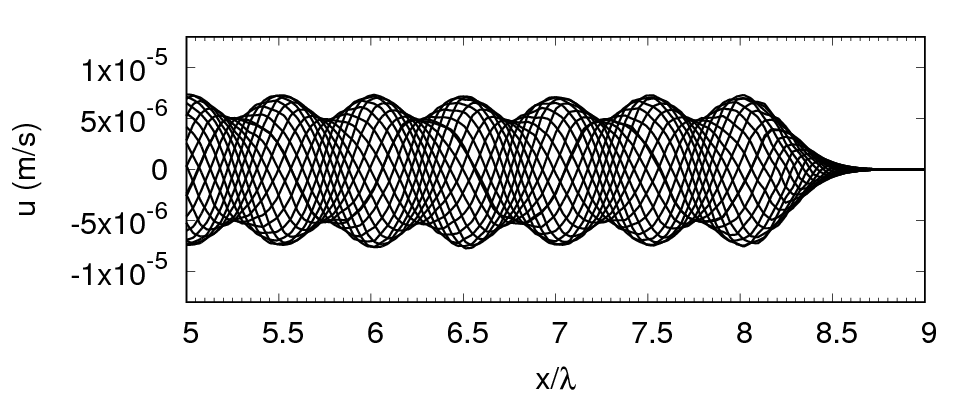}
\end{center}
\caption{Theory predictions (left) and simulation results (right) for velocity in water over $x$-coordinate, given at equally spaced time instances during the last simulated period; for forcing strengths $\gamma = 2560\, \mathrm{s^{-1}}$ (top), $\gamma = 10240\, \mathrm{s^{-1}}$ (middle), $\gamma = 81920\, \mathrm{s^{-1}}$ (bottom); for exponential blending via Eq. (\ref{EQblendEXP}) and $x_{\mathrm{d}} = 1\lambda$} \label{FIGw1expvel}
\end{figure}

\begin{figure}[H]
\begin{center}
\includegraphics[width=\halffactor\linewidth]{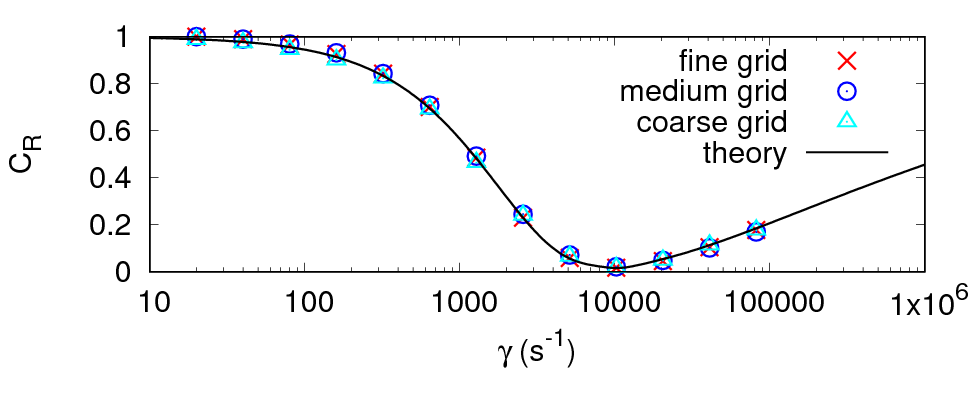}
\includegraphics[width=\halffactor\linewidth]{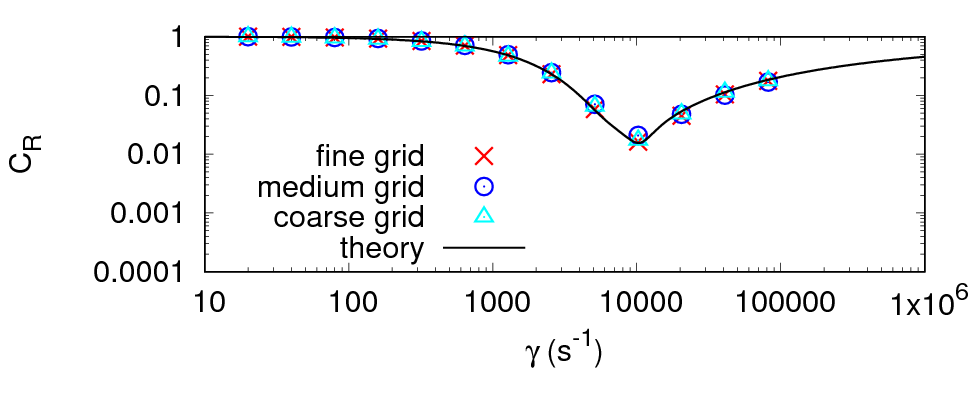}
\end{center}
\caption{Reflection coefficient $C_{\mathrm{R}}$ over forcing strength $\gamma$  from simulation and theory;  for exponential blending via Eq. (\ref{EQblendEXP}), layer thickness $x_{\mathrm{d}} = 1\lambda$, wavelength $\lambda$ and wave period $T$ as in Fig. \ref{FIGw1expvel}; for coarse (time step $\Delta t = T/100$, cell size $\Delta x = \lambda/30$), medium ($\Delta t = T/200$, $\Delta x = \lambda/60$), and fine ($\Delta t = T/400$, $\Delta x = \lambda/120$) discretization } \label{FIGw1expCr}
\end{figure}

Figures \ref{FIGw1expvel} and \ref{FIGw1expCr} show velocities and reflection coefficients for sound waves in water. The layer thickness is $x_{\mathrm{d}}=1\lambda$ and exponential blending according to Eq. (\ref{EQblendEXP}) is used. Theory and simulation results agree well.
The peaks in the partial standing wave in Fig. \ref{FIGw1expvel} have different locations depending on forcing strength $\gamma$. This shows that, with increasing $\gamma$, the effective reflection location shifts from the boundary, to which the layer is attached (here $x/\lambda=9$), towards the entrance to the absorbing layer (here $x/\lambda=8$); this underlines the importance of including reflections which occur within the layer in the analysis.

\begin{figure}[H]
\begin{center}
\includegraphics[width=\halffactor\linewidth]{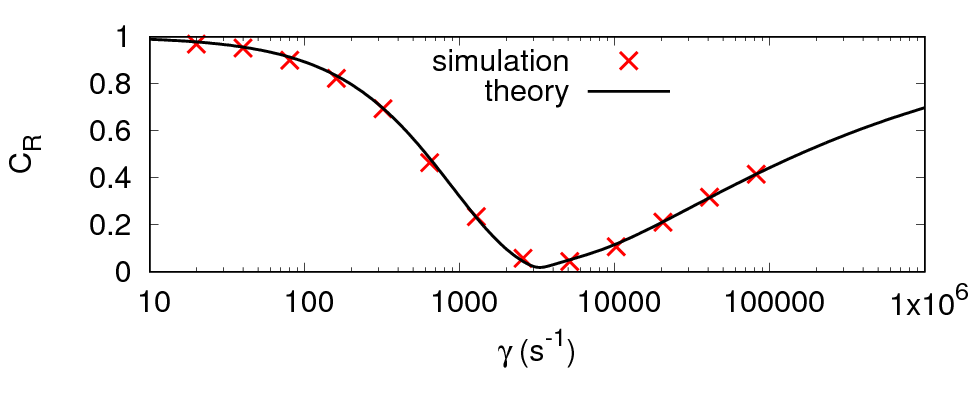}
\includegraphics[width=\halffactor\linewidth]{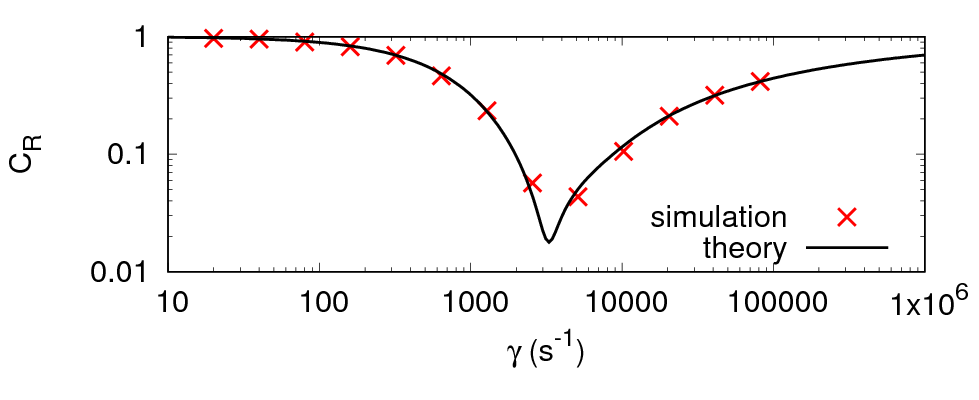}
\end{center}
\caption{Reflection coefficient $C_{\mathrm{R}}$ over forcing strength $\gamma$ from simulation and theory; for linear blending via Eq. (\ref{EQblendLIN}) and $x_{\mathrm{d}} = 1\lambda$} \label{FIGw1linCr}
\end{figure}

\begin{figure}[H]
\begin{center}
\includegraphics[width=\halffactor\linewidth]{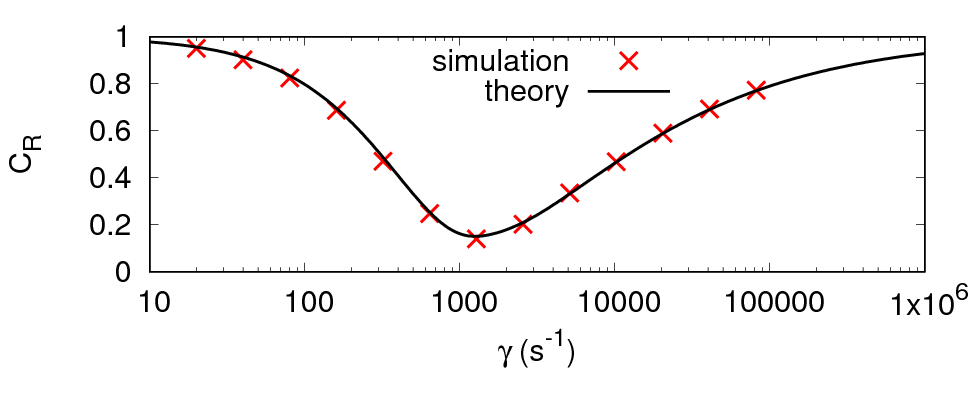}
\includegraphics[width=\halffactor\linewidth]{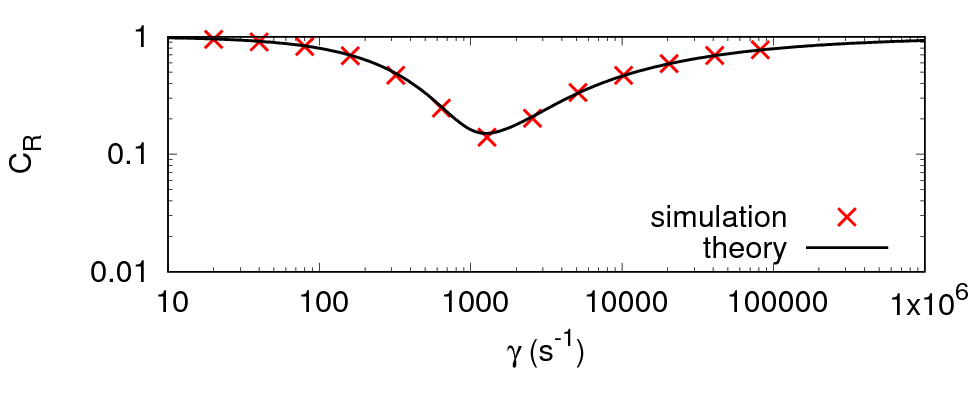}
\end{center}
\caption{Reflection coefficient $C_{\mathrm{R}}$ over forcing strength $\gamma$ from simulation and theory; for constant blending via Eq. (\ref{EQblendCONST}) and $x_{\mathrm{d}} = 1\lambda$ } \label{FIGw1constCr}
\end{figure}

In Fig. \ref{FIGw1expCr}, results are given for the grid and time step from Sect. \ref{SECsimsetup}, and also for twice and four times refined mesh and time step size. The difference between the results are small, thus the coarsest discretization is used for the rest of the simulations in this section.  
Subsequently, the simulations are rerun with same setup, except once with linear blending (Fig. \ref{FIGw1linCr}) and once with constant blending (Fig. \ref{FIGw1constCr}). Again the results show good agreement between theory and simulation.  Compared to Figs. \ref{FIGw1expvel} and \ref{FIGw1expCr}, the optimum values $\gamma_{\mathrm{opt}}$ of the forcing strength are different. For the investigated blending functions it holds roughly that the smaller the area below the blending function $b(x)$ is, the larger is the value for $\gamma_{\mathrm{opt}}$.
The results confirm that higher order blending functions should be preferred.

\begin{figure}[H]
\begin{center}
\includegraphics[width=\halffactor\linewidth]{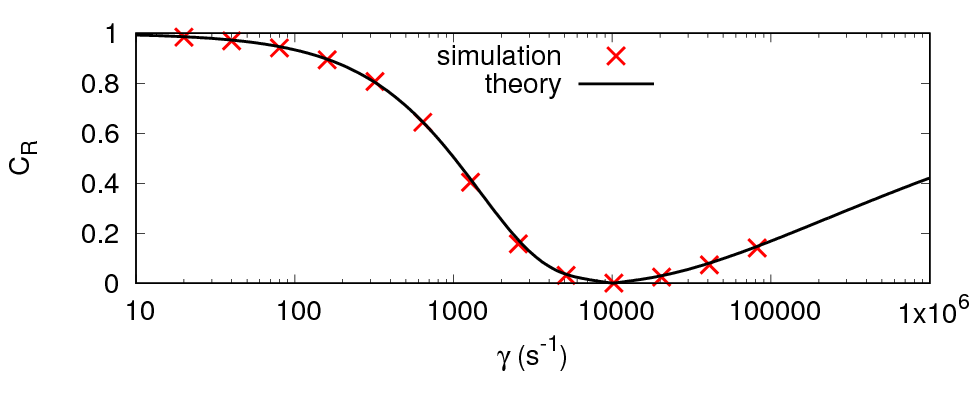}
\includegraphics[width=\halffactor\linewidth]{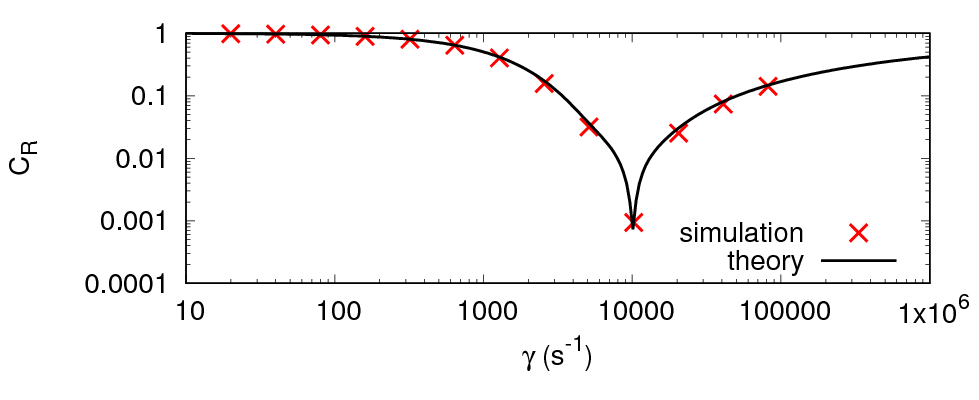}
\end{center}
\caption{Reflection coefficient $C_{\mathrm{R}}$ over forcing strength $\gamma$ from simulation and theory; for exponential blending via Eq. (\ref{EQblendEXP}) and $x_{\mathrm{d}} = 1.18\lambda$} \label{FIGa1p2expCr}
\end{figure}

\begin{figure}[H]
\begin{center}
\includegraphics[width=\halffactor\linewidth]{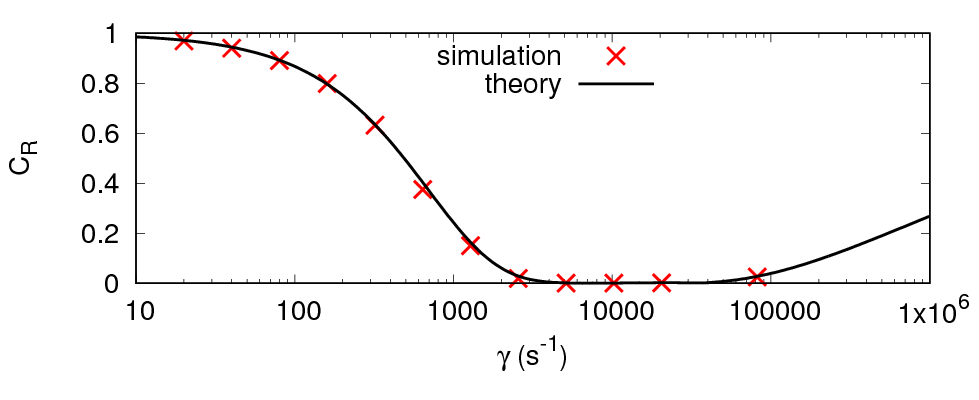}
\includegraphics[width=\halffactor\linewidth]{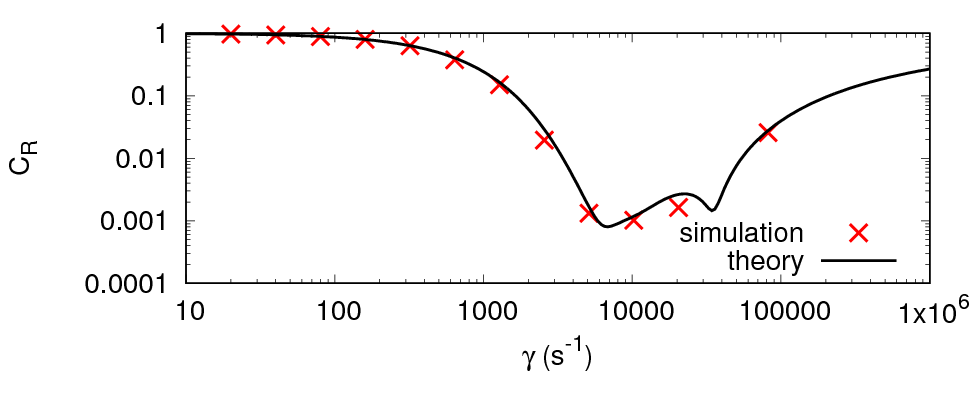}
\end{center}
\caption{Reflection coefficient $C_{\mathrm{R}}$ over forcing strength $\gamma$ from simulation and theory;  for exponential blending via Eq. (\ref{EQblendEXP}) and $x_{\mathrm{d}} = 2.35\lambda$} \label{FIGa2expCr}
\end{figure}

Simulation results for acoustic waves in an ideal gas are shown in Figs. \ref{FIGa1p2expCr} to \ref{FIGa2expCr}  for  absorbing layers with thickness $x_{\mathrm{d}} = 1.18\lambda$ and $x_{\mathrm{d}} = 2.35\lambda$. Theory predictions and simulation results are in good agreement.
Figure \ref{FIGa1p2expCr} (at $\gamma = 10 240\, \mathrm{s}^{-1}$) demonstrates that, for certain choices of $x_{\mathrm{d}}$ and for a very narrow range of wave frequencies, it is possible to achieve reflection coefficients more than one order of magnitude smaller than are possible for slightly smaller or larger $x_{\mathrm{d}}$. This is expected to be an effect of especially favorable destructive interference between the reflected waves.
Figure \ref{FIGa2expCr} shows that there may be more than one local optimum for $\gamma$. This should be taken into account in future research, e.g. when searching for an optimum formulation for $b(x)$.

The last 1D example is the prediction of reflection coefficients for irregular wave trains. Apart from the following modifications, the setup is the same as in the previous simulations. The fluid is air as ideal gas with speed of sound $c=347.28\, \mathrm{\frac{m}{s}}$ and the termperature variation is accounted for via Eq. (\ref{EQenergy}). The domain length is $L_{\mathrm{x}}=13.36\lambda_{\mathrm{peak}}$, where $\lambda_{\mathrm{peak}}\approx 0.78\, \mathrm{m}$ is the wavelength corresponding to peak wave period $T_{\mathrm{peak}}$. The wave absorbing layer based on Eq. (\ref{EQmomdamp}) is attached to boundary  $x=L_{\mathrm{x}}$, with zone thickness $x_{\mathrm{d}}=2\lambda_{\mathrm{peak}}$ and exponential blending as in Eq. (\ref{EQblendEXP}).
To generate irregular waves, pressure fluctuations at the inlet are prescribed using the JONSWAP-spectrum by Hasselmann et al. (1973), except that the 'significant wave height' $H_{\mathrm{s}}$ was replaced by the twice the 'significant pressure amplitude' $2p_{\mathrm{a,s}}=10\, \mathrm{Pa}$; the parameters are peak wave period $T_{\mathrm{peak}}=0.002\overline{27}\, \mathrm{s}$, and a peak-shape parameter of $3.3$. The spectrum is discretized into $50$ wave components.

The waves are generated  by prescribing the pressure at the inlet as $p(t)b'(t)$, where $p(t)$ corresponds to a linear superposition of the pressures of the irregular wave components, and  $b'(t)$ is  a blending-in and -out during the first $6.6T_{\mathrm{peak}}$ of simulation time:
\begin{equation}
b'(t)=
\begin{cases}
    \cos^{2}  \left( \frac{\pi}{2} (t - 2.2T_{\mathrm{peak}})/(2.2T_{\mathrm{peak}}) \right)      & \quad \text{if } t<2.2T_{\mathrm{peak}}\\
    1.0  & \quad \text{if } 2.2T_{\mathrm{peak}} \leq t\leq 4.4T_{\mathrm{peak}}\\
    \cos^{2} \left( \frac{\pi}{2} (t - 4.4T_{\mathrm{peak}})/(2.2T_{\mathrm{peak}})\right)       & \quad \text{if } 4.4T_{\mathrm{peak}}<t<6.6T_{\mathrm{peak}}\\
    0.0 & \quad \text{if } t\geq 6.6T_{\mathrm{peak}}
  \end{cases} \quad .
\label{EQcos2fade}
\end{equation}

\begin{figure}[H]
\begin{center}
\includegraphics[width=\halffactor\linewidth]{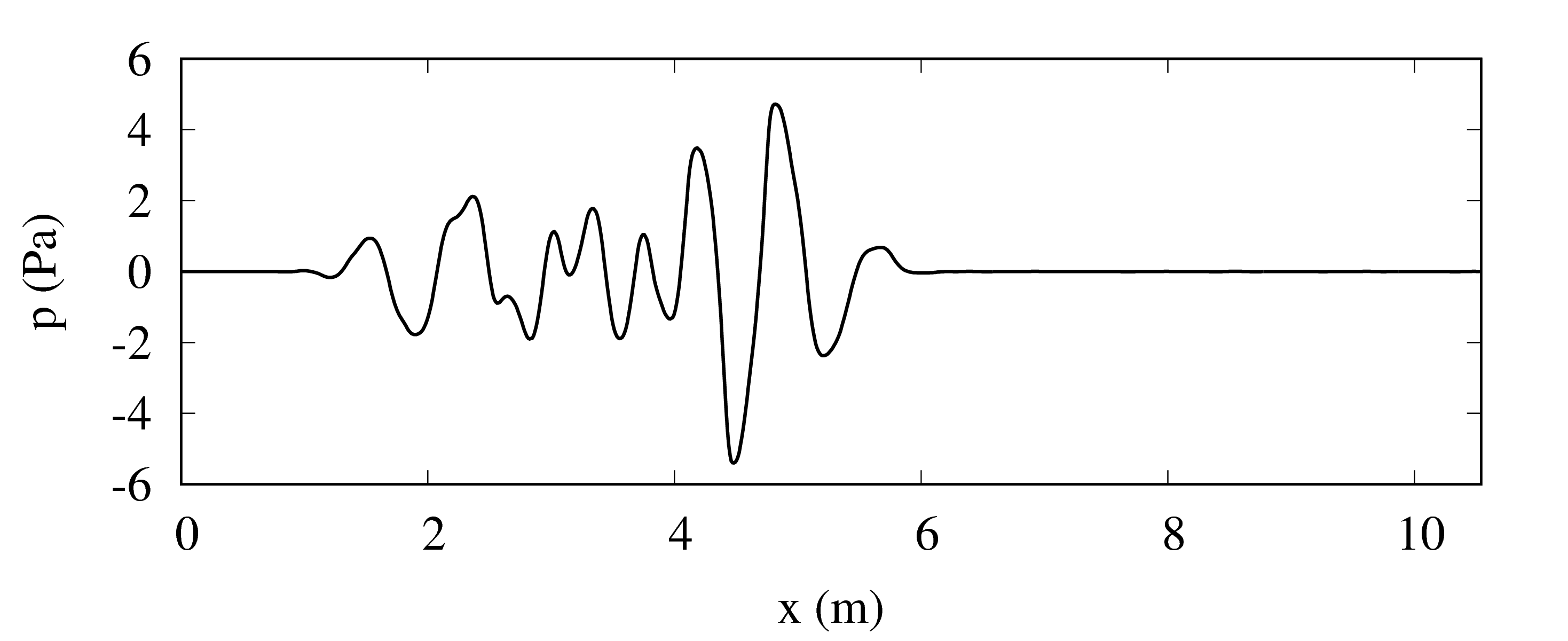}
\end{center}
\caption{Pressure in domain at time $t=21.27T_{\mathrm{peak}}$ with no damping ($\gamma=0$)} \label{FIGPoXnodamp}
\end{figure}

The total simulated time is $t_{\mathrm{end}}=21.27T_{\mathrm{peak}}$, and Fig. \ref{FIGPoXnodamp} shows the pressure in the domain at this time if no damping is applied. The waves are discretized by $\approx 62$ cells per peak wavelength $\lambda_{\mathrm{peak}}$ and the time step is $ T_{\mathrm{peak}}/200$, with $10$  iterations  per time step. 
At the end of each simulation, the pressure over space is analysed using the Fast-Fourier Transform (FFT) algorithm to obtain the pressure amplitude spectrum of the reflected wave. For $\gamma=0$, the wave is perfectly reflected, so the resulting  pressure amplitude spectrum corresponds to the initially generated spectrum; thus for $\gamma \neq 0$, the amount of reflection can be judged by comparing the pressure amplitude spectrum for the reflected wave to the case for no damping ($\gamma=0$). The global reflection coefficient $C_{\mathrm{R}}$ was obtained by Eq. (\ref{EQCrB2}).

\begin{figure}[h]
\includegraphics[width=\halffactor\linewidth]{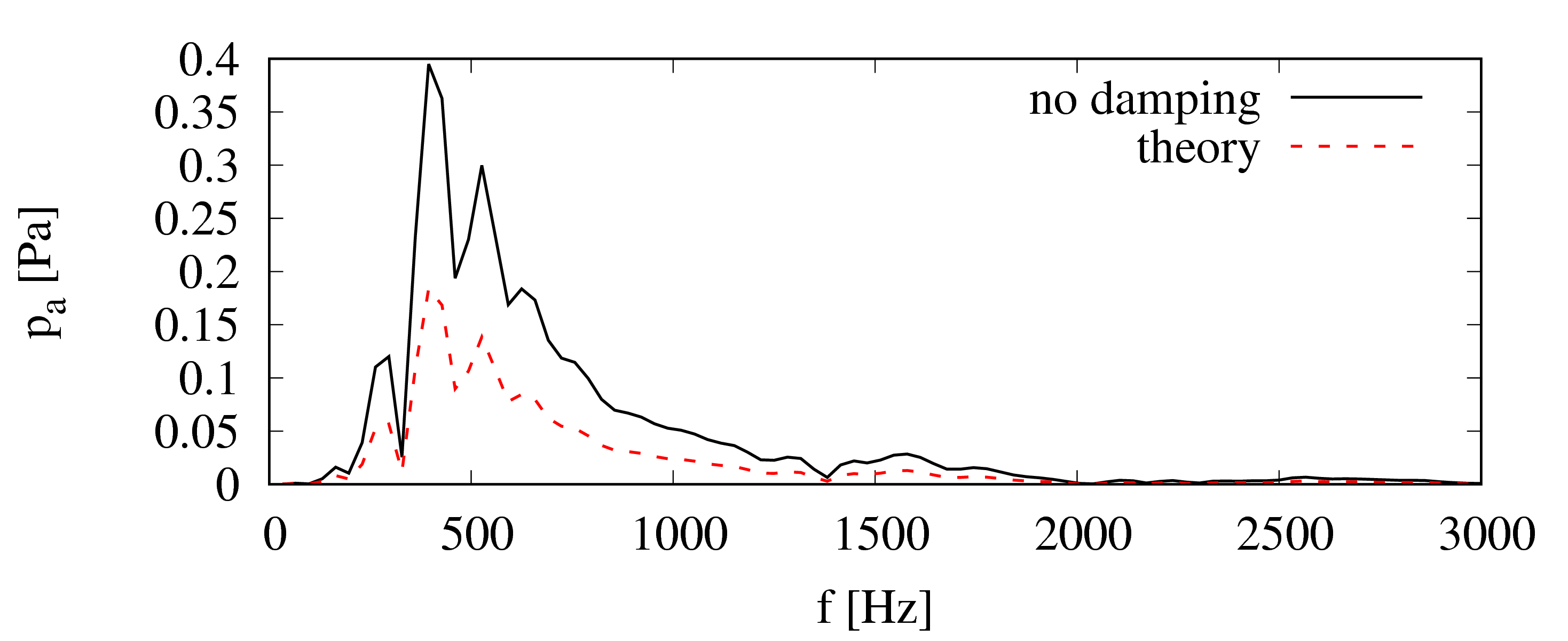}\includegraphics[width=\halffactor\linewidth]{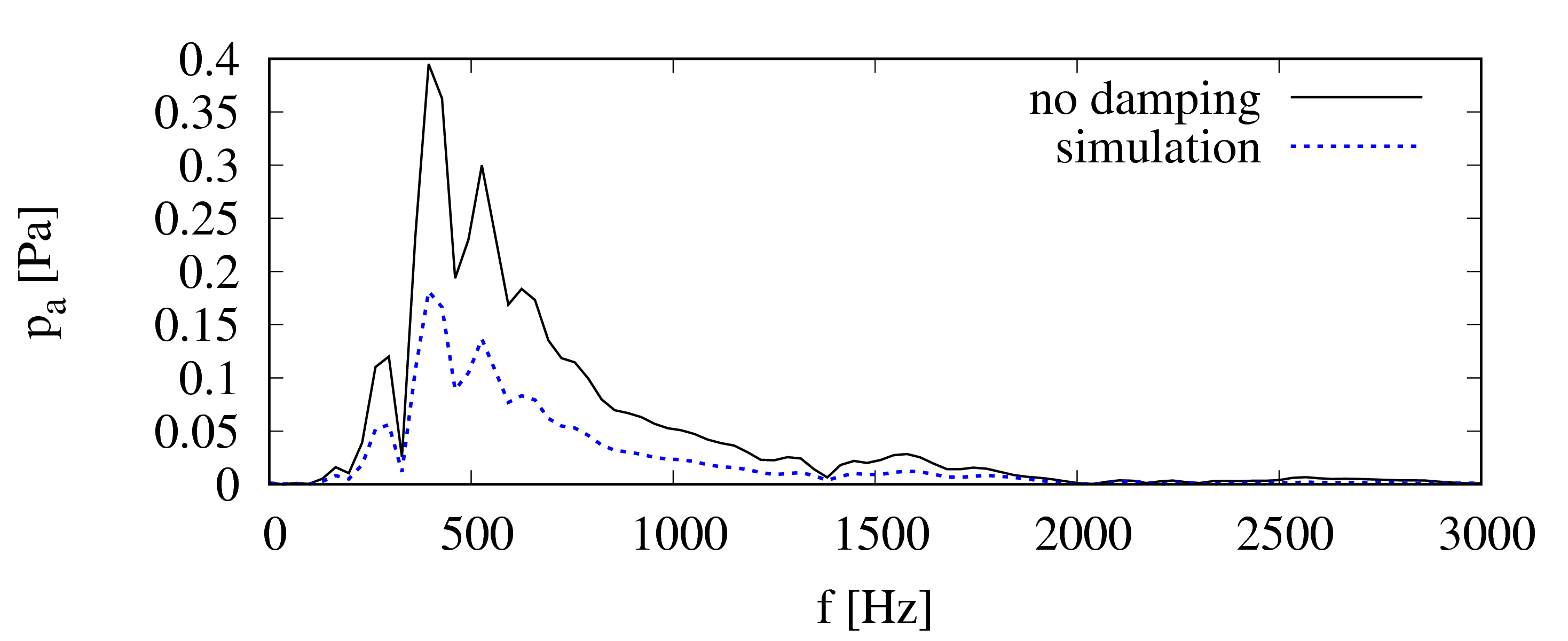}\\
\includegraphics[width=\halffactor\linewidth]{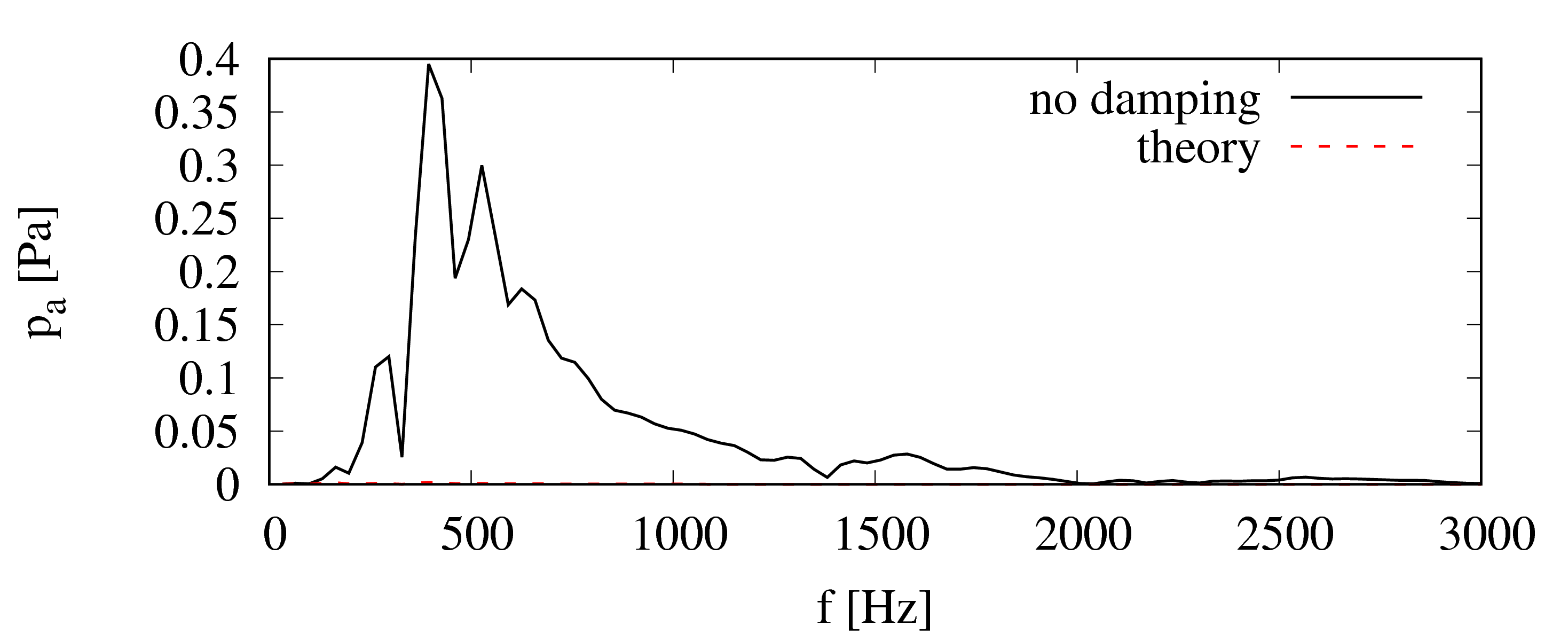}\includegraphics[width=\halffactor\linewidth]{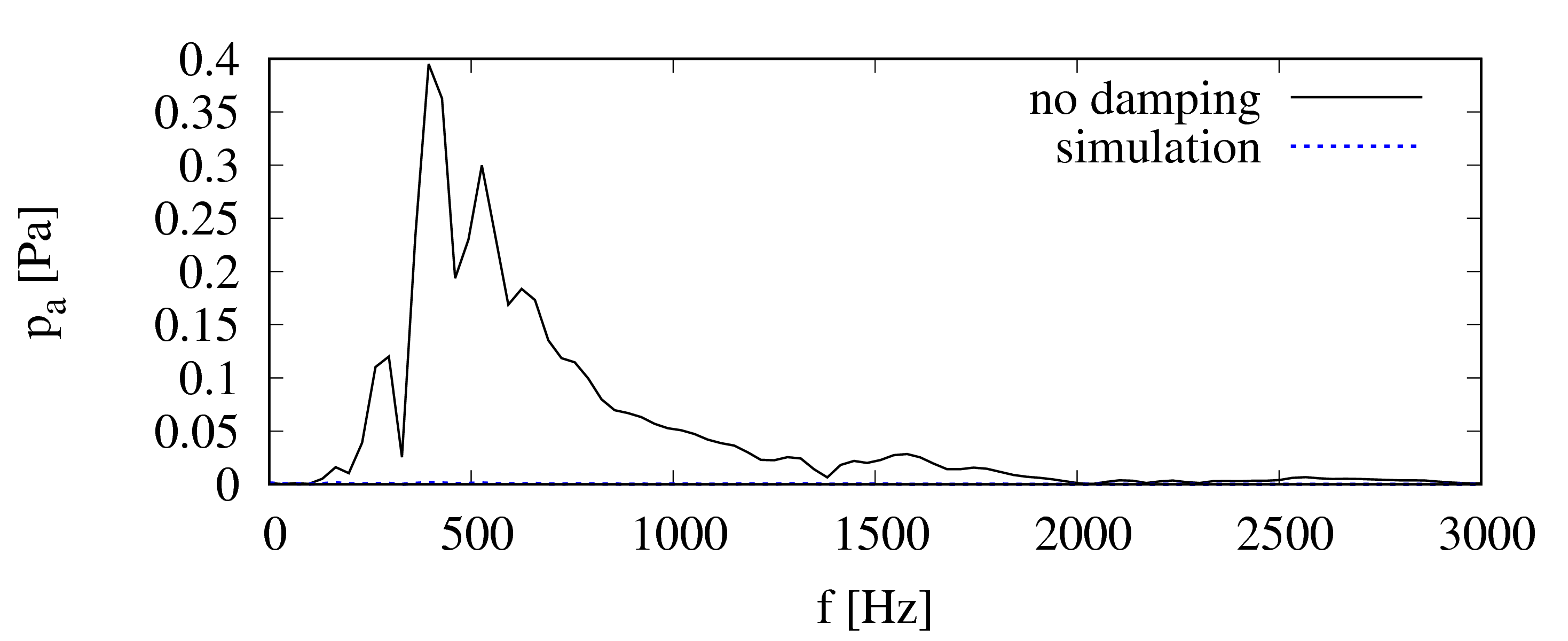}\\
\includegraphics[width=\halffactor\linewidth]{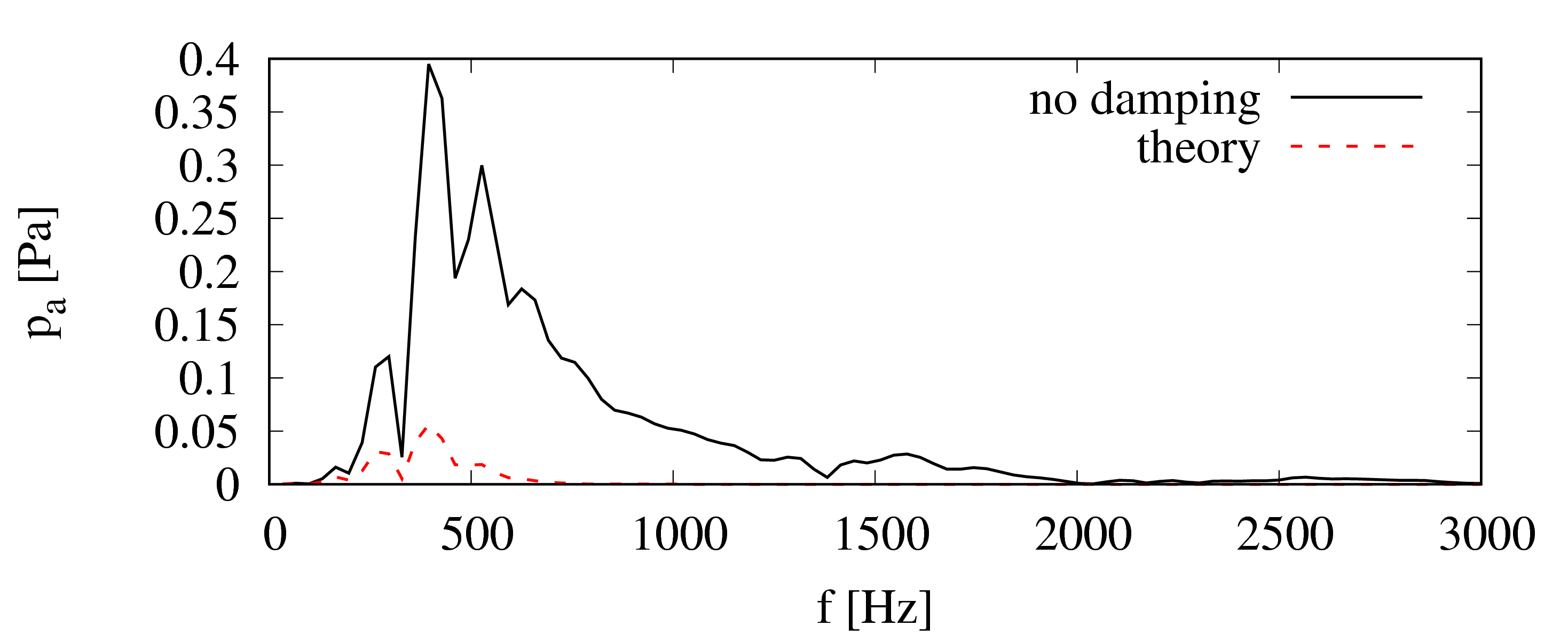}\includegraphics[width=\halffactor\linewidth]{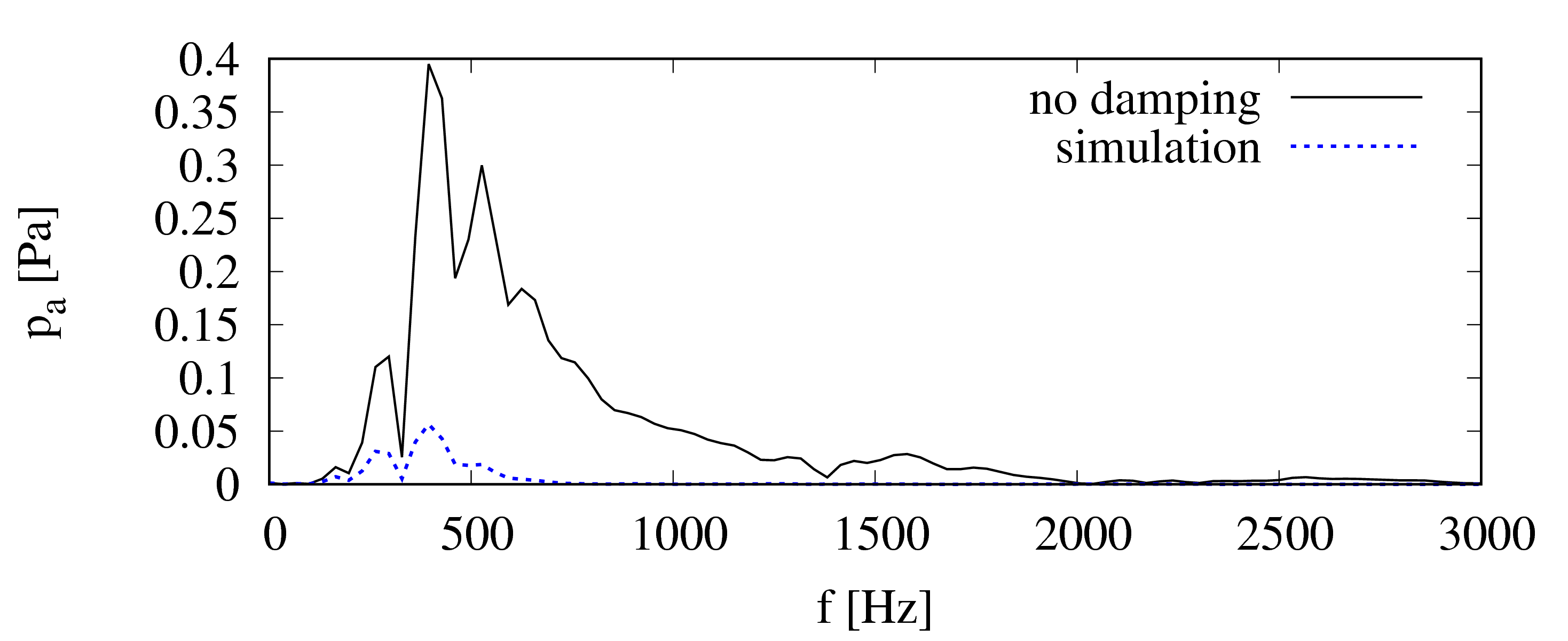}
\caption{Theory predictions (left) and simulation results (right) for generated and reflected pressure-amplitude spectrum over wave frequency; for too weak damping (top image, $\gamma=640\, \mathrm{s}^{-1}$), roughly optimum damping (middle image, $\gamma=5120\, \mathrm{s}^{-1}$), and too strong damping (bottom image, $\gamma=163840\, \mathrm{s}^{-1}$) } \label{FIGCr4waves}
\end{figure}

Figure \ref{FIGCr4waves} shows that the resulting pressure amplitude spectrums agree well with the predictions according to the theory from  Sect. \ref{SECtheory}; the theory predictions are based on the assumption that the wave can be treated as a linear superposition of wave components with different frequencies, and that the reflection coefficient for each component equals the reflection coefficient as predicted by the theory from  Sect. \ref{SECtheory}.

\begin{figure}[h]
\includegraphics[width=\halffactor\linewidth]{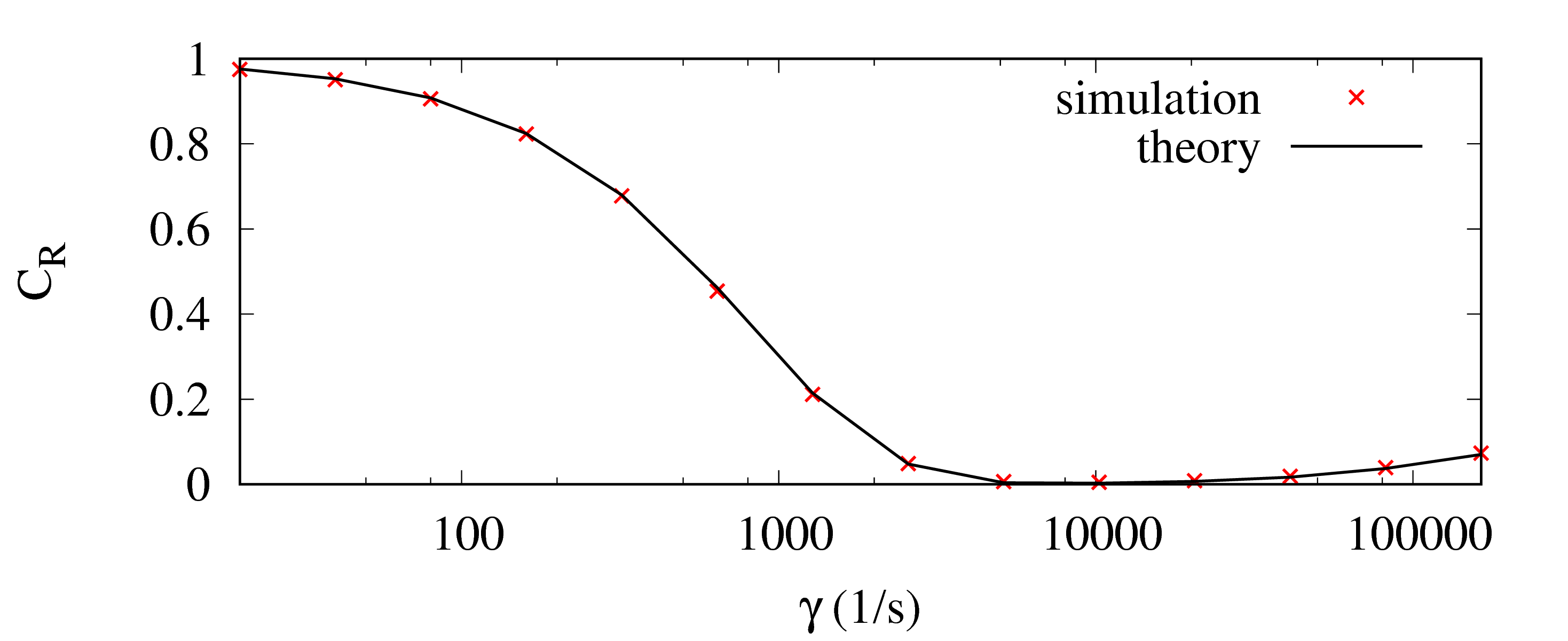}
\includegraphics[width=\halffactor\linewidth]{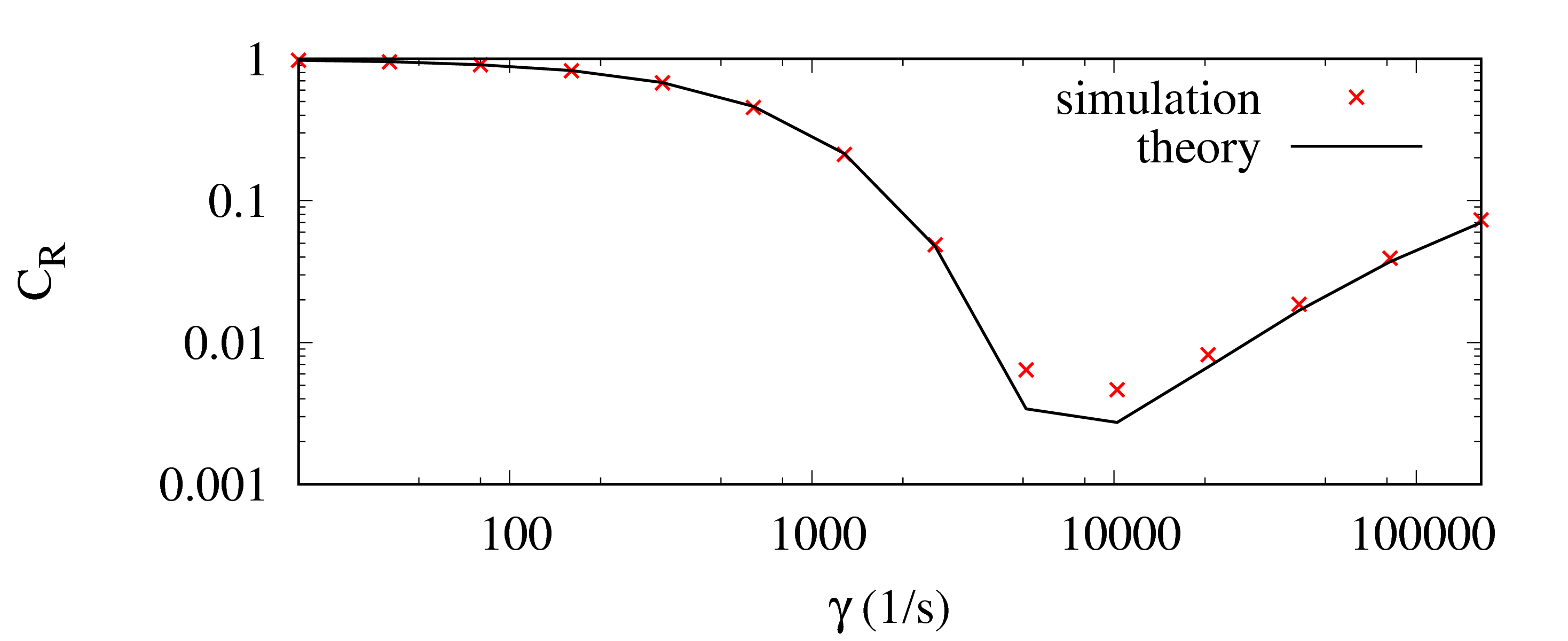}
\caption{Reflection coefficient $C_{\mathrm{R}}$ according to Eq. (\ref{EQCrB2}) over forcing strength $\gamma$, for simulation results and theory prediction; the right image corresponds to the left image with a logarithmic $y$-axis } \label{FIGspecCRcomparesimtheory}
\end{figure}

Figure \ref{FIGspecCRcomparesimtheory} shows that the resulting overall reflection coefficients agree well for simulation and theory; the differences in $C_{\mathrm{R}}$ between simulation and theory are $<1\%$.

\section{Discussion Based on Theory Results}
\label{SECDiscusstheory}
Since Sect. \ref{SECres} demonstrated that the theory is reliable in its predictions, this section shows theory results without backup from simulation results. This is mainly because, to obtain the following results,  a very large  number of simulations would be required, the combined computational effort being out of the scope of this study. 

\subsection{Theory implications for the choice of blending function $b(x)$}
\label{SECDiscusstheoryBlend}
\begin{figure}[H]
\begin{center}
(a)\includegraphics[width=0.45\linewidth]{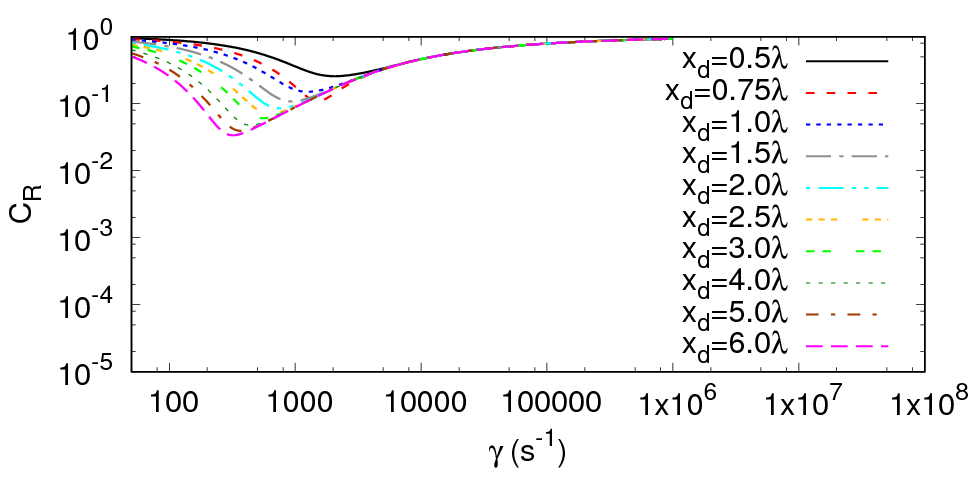}
(b)\includegraphics[width=0.45\linewidth]{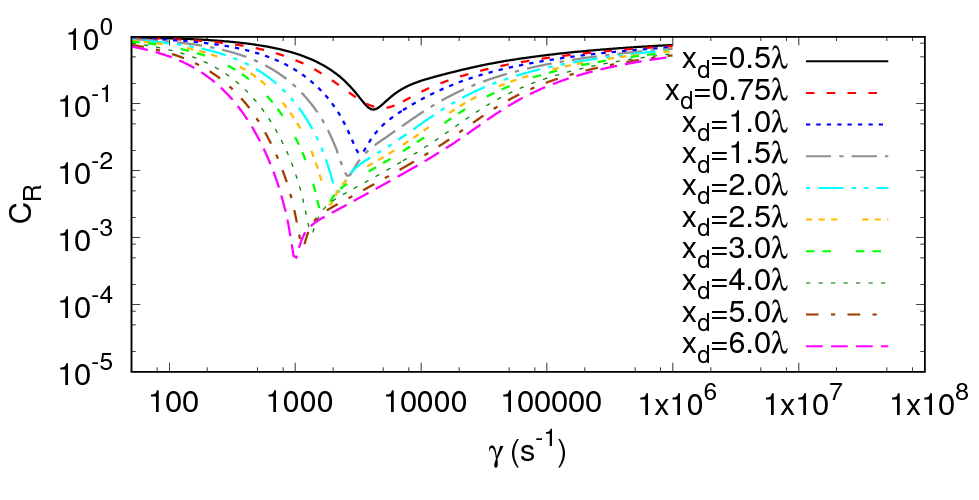}\\
(c)\includegraphics[width=0.45\linewidth]{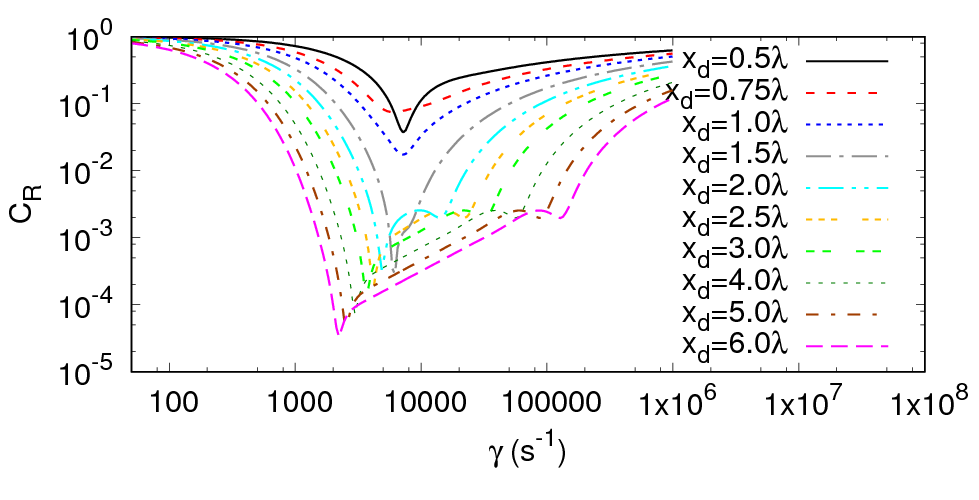}
(d)\includegraphics[width=0.45\linewidth]{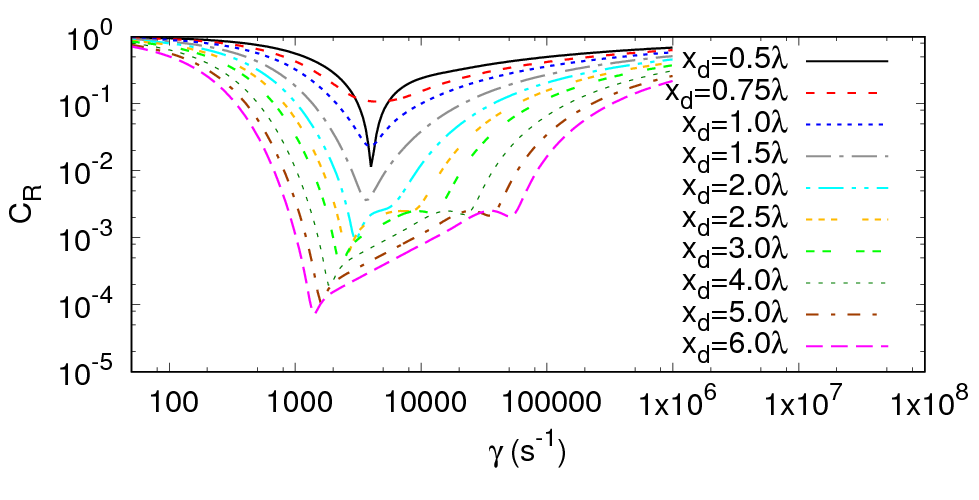}\\
(e)\includegraphics[width=0.45\linewidth]{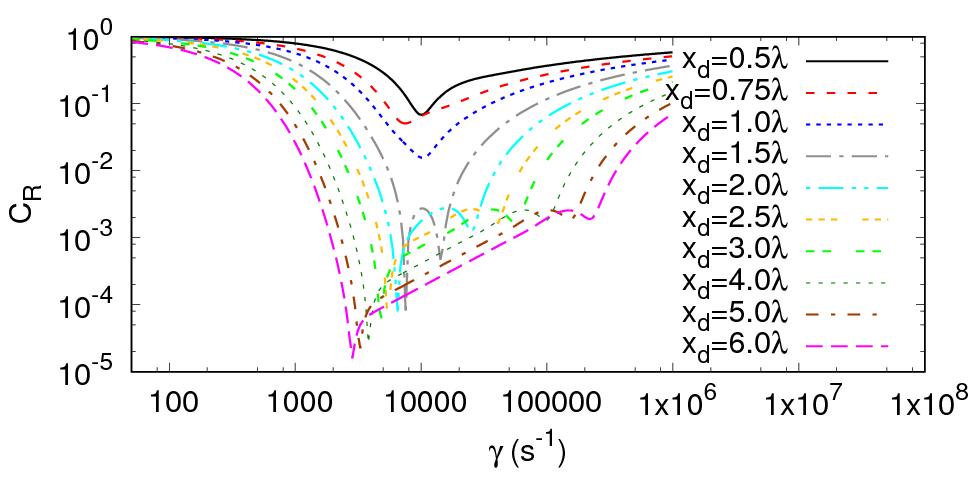}
\end{center}
\caption{Reflection coefficient $C_{\mathrm{R}}$ over forcing strength $\gamma$ for different absorbing layer thicknesses $x_{\mathrm{d}}$; for constant (a), linear (b), quadratic (c), cosine-squared (d) and exponential (e) blending as given in Sect. \ref{SECabslayer}} \label{FIGtheorydiffblend}
\end{figure}

Although this work was not intended to fully answer which is the optimum choice for blending function $b(x)$, it can guide regarding the choice between the $b(x)$ from Sect. \ref{SECabslayer}. For this, the theory is evaluated for a wave with period $T=2.\overline{27}\cdot 10^{-3}\, \mathrm{s}$ and absorbing layer thicknesses varying between $x_{\mathrm{d}} \in [0.5\lambda,6\lambda]$. These results can easily be applied to waves of any other period: If $\gamma$ and $x_{\mathrm{d}}$ are scaled as described in Sect. \ref{SECabslayer}, the curves in Fig. \ref{FIGtheorydiffblend} need only to be shifted sideways accordingly.

As Fig. \ref{FIGtheorydiffblend} shows, the higher-order blending functions give better results than linear or even constant $b(x)$. Although the curves for quadratic, cosine-squared and exponential blending are substantially different as seen in Fig. \ref{FIGblend}, their results seem on the whole nearly equally satisfactory, perhaps with a slight preference for the exponential blending function. 

\begin{figure}[H]
\begin{center}
(a)\includegraphics[width=0.5\linewidth]{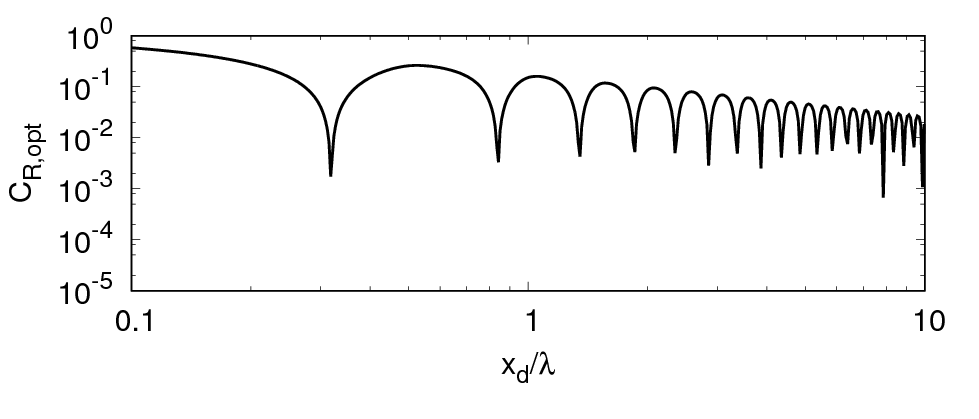}\includegraphics[width=0.5\linewidth]{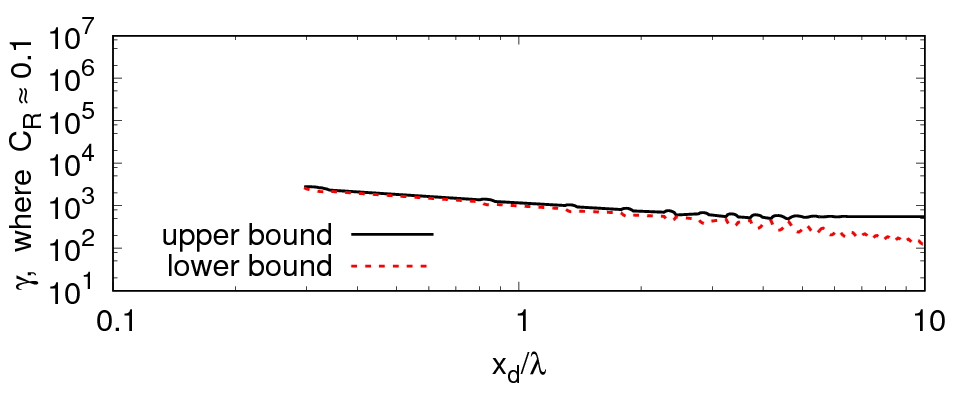}\\
(b)\includegraphics[width=0.5\linewidth]{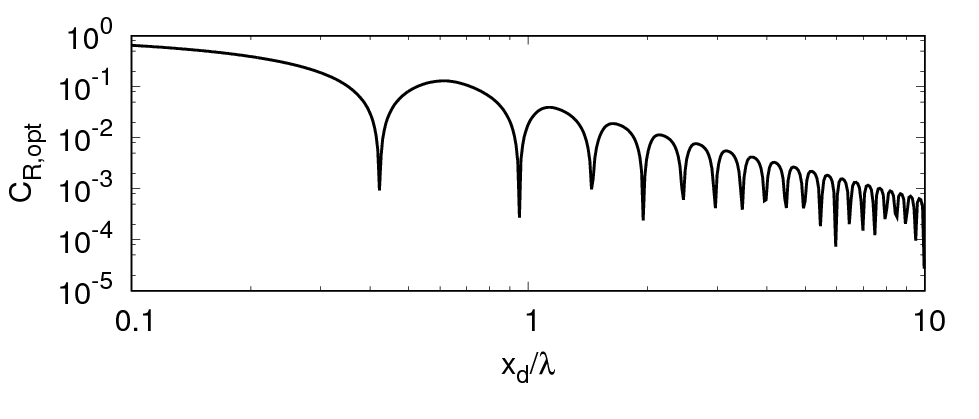}\includegraphics[width=0.5\linewidth]{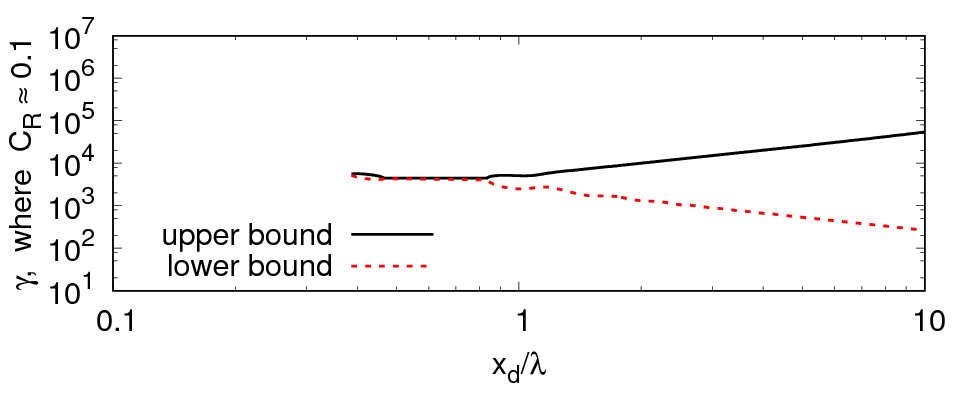}\\
(c)\includegraphics[width=0.5\linewidth]{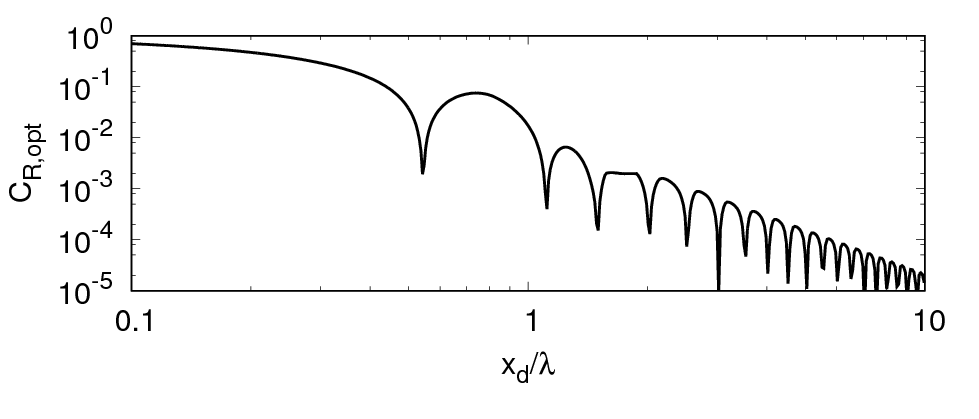}\includegraphics[width=0.5\linewidth]{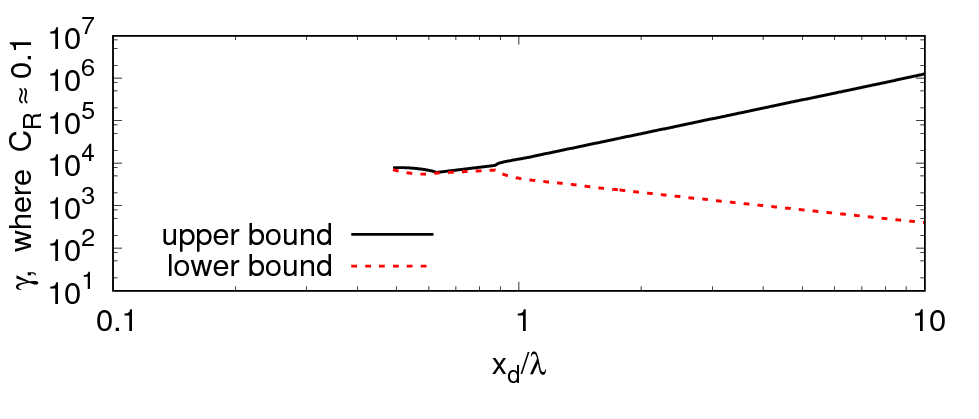}\\
(d)\includegraphics[width=0.5\linewidth]{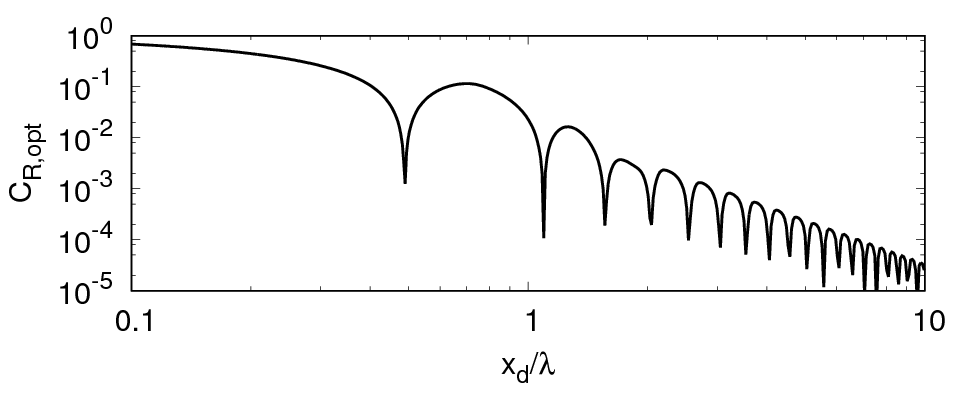}\includegraphics[width=0.5\linewidth]{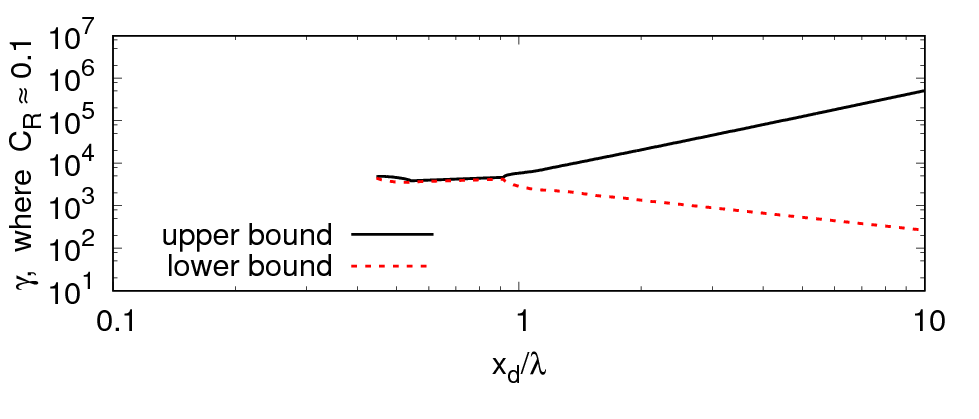}\\
(e)\includegraphics[width=0.5\linewidth]{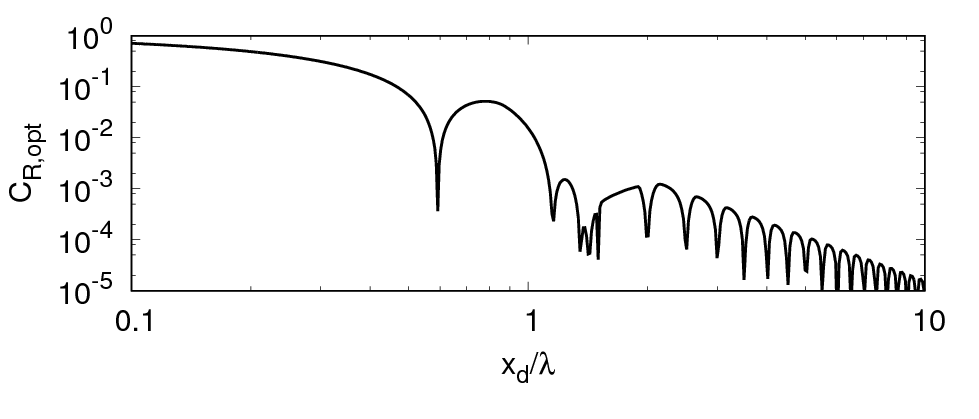}\includegraphics[width=0.5\linewidth]{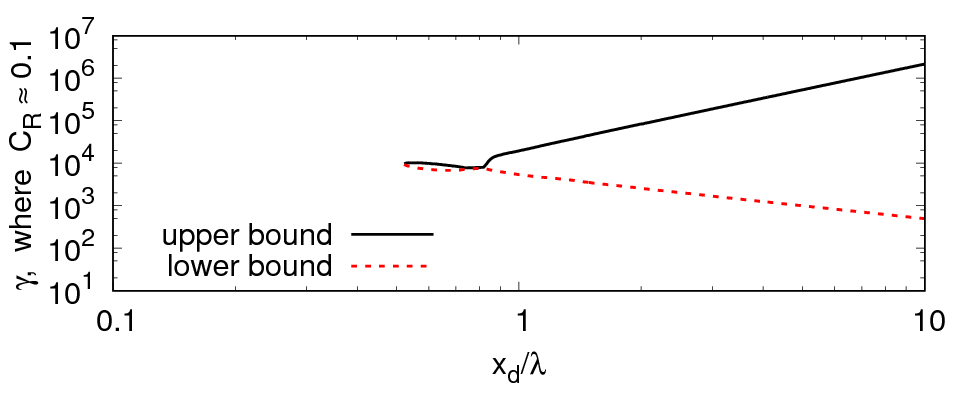}
\end{center}
\caption{Optimum reflection coefficient $C_{\mathrm{R,opt}}$ (left) and range of $\gamma$ values for which $C_{\mathrm{R}}<0.1$ (right) over layer thickness $x_{\mathrm{d}}$ per wavelength $\lambda$ for constant (a), linear (b), quadratic (c), cosine-squared (d) and exponential (e) blending as given in Sect. \ref{SECabslayer}} \label{FIGtheoryrangecr}
\end{figure}

For some values of $x_{\mathrm{d}}$ in Fig. \ref{FIGtheorydiffblend}, quadratic or cosine-squared blending produces lower optimum reflection coefficients than exponential blending, e.g. for $x_{\mathrm{d}}=0.5\lambda$. 
This can be explained by Fig. \ref{FIGtheoryrangecr}, which shows that there exist certain  values for absorbing layer thickness $x_{\mathrm{d}}$, for which especially favorable absorption may occur. This means
that the resulting reflection coefficient $C_{\mathrm{R}}$ for optimum forcing strength $\gamma$ can be two orders of magnitude lower than it would be if  $x_{\mathrm{d}}$ was selected slightly larger or smaller. 
This phenomenon may be used to produce satisfactory wave damping with very thin absorbing layers. The results in Fig. \ref{FIGa1p2expCr} confirm the existence of such a phenomenon. It is expected that a combination of grid stretching and absorbing layers may produce similar effects.

However, according to the present investigations, these settings, seen as 'negative peaks' in Fig. \ref{FIGtheoryrangecr}, occur only for a very narrow range of wavelengths. 
This would make the absorption behavior  rather sensitive to slight deviations in the wavelength. For example in finite-volume-based flow solvers, discretization and iteration errors can lead to a change in the wavelength of several percent for coarse discretizations\cite{REFperic2016}. Furthermore, when a superposition of waves of different frequencies is considered, then it is more important to have satisfactory minimization of all undesired reflections, not only of those belonging to a specific wave component.
Trying to optimize the absorbing layer towards one of these negative peaks of $C_{\mathrm{R}}$ may therefore lead to false security, and should only be exercised with caution.

In practice, it is desired (especially when the wave may be modified within the domain) to know for a given absorbing layer setup the range of forcing strengths $\gamma$ for which the reflection coefficients will be below a certain threshold. As shown for an example threshold of $C_{\mathrm{R}}<10\%$ in Fig. \ref{FIGtheoryrangecr}, this range is larger for the higher-order blending functions.

\subsection{Convergence of theory to solution for continuous blending}
\label{SECDiscusstheoryconvergence}

\begin{figure}[H]
\begin{center}
\includegraphics[width=\halffactor\linewidth]{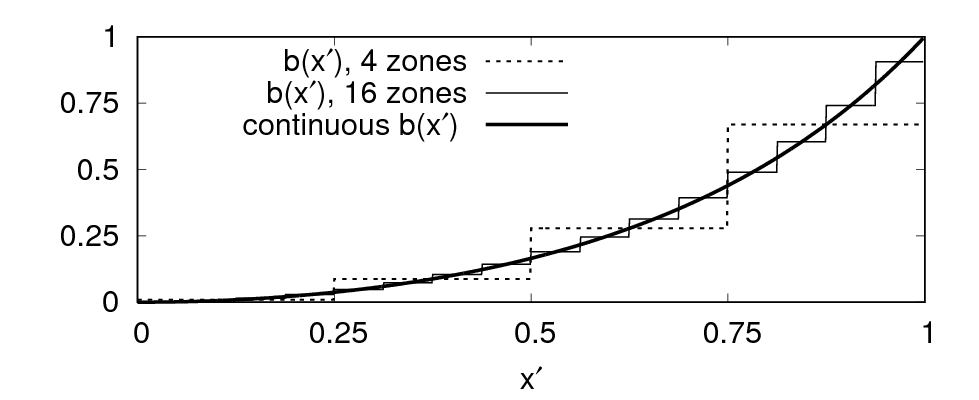} 
\end{center}
\caption{Blending function $b(x')$ according to Eq. (\ref{EQblendEXP}) over $x'$-coordinate; $x'$ is directed in wave propagation direction and linearly scaled such that it is $0$ at the entrance to the absorbing layer and $1$ at the boundary to which the zone is attached; for absorbing layer consisting of $4$ and $16$ zones and for continuous $b(x')$ } \label{FIGdiscBcomp}
\end{figure}

The theory from Sect. \ref{SECtheory} subdivides the absorbing layer into $n$ zones with constant blending $b(x)$ as illustrated in Fig. \ref{FIGdiscBcomp}. 
This section demonstrates that, if $n$ is larger than a certain threshold, then the theory results can be considered independent of $n$. Thus also the wave damping in flow simulations is basically grid-independent, if the number $n$ of grid cells, by which the absorbing layer is discretized in wave propagation direction, is above the same threshold.

\begin{figure}[H]
\begin{center}
\includegraphics[width=\halffactor\linewidth]{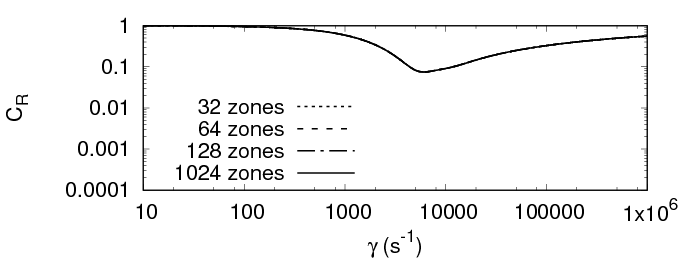}
\includegraphics[width=\halffactor\linewidth]{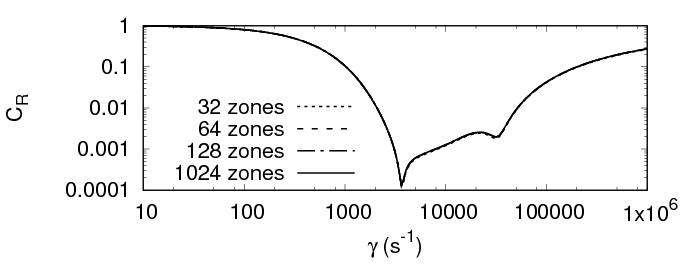}
\end{center}
\caption{Theory prediction for reflection coefficient $C_{\mathrm{R}}$ over forcing strength $\gamma$ for waves with period $T=0.002\overline{27}\, \mathrm{s}$; for an absorbing layer according to Eq. (\ref{EQmomdampTheory}) with exponential blending  (Eq. (\ref{EQblendEXP})), zone thickness $x_{\mathrm{d}} = 0.75\lambda $ (top),  and $x_{\mathrm{d}} = 3.0\lambda $ (bottom);  for   $b(x)$ subdivided into $32$, $64$, $128$ and $1024$ zones } \label{FIGCr0gdisc}
\end{figure}

Figure \ref{FIGCr0gdisc}  shows for a subdivision into $32$ zones, that the results are barely distinguishable from subdivisions into larger numbers of zones. This agrees well with findings from Sect. \ref{SECres}, where the wave damping was observed to be grid-independent for practical flow simulation setups (i.e. grids with at least $30$ cells per wavelength). 

\begin{figure}[H]
\begin{center}
\includegraphics[width=\halffactor\linewidth]{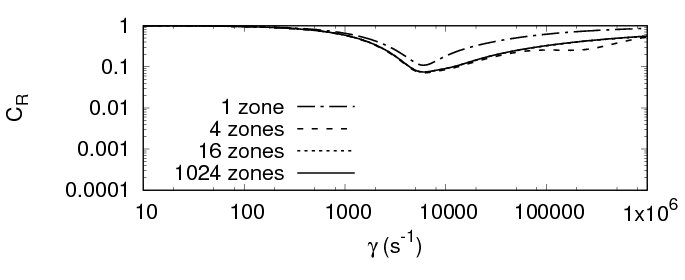} 
\includegraphics[width=\halffactor\linewidth]{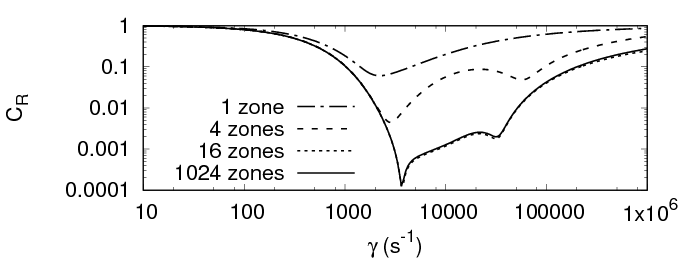} 
\end{center}
\caption{As Fig. \ref{FIGCr0gdisc}, except that $b(x)$ is subdivided into $1$, $4$, $16$ and $1024$ zones} \label{FIGCr0gdiscLOG}
\end{figure}

Figure \ref{FIGCr0gdiscLOG} shows that for subdivision into less than $16$ zones, the results differ significantly from the results for $>32$ zones. This is relevant when assessing flow simulations, in which a combination of  grid stretching and absorbing layers is used to damp the waves; in industrial practice, these two wave damping approaches are sometimes combined with the intention to lower the computational effort and to improve the damping. However, Figs. \ref{FIGCr0gdisc} and \ref{FIGCr0gdiscLOG} show that, if due to the grid stretching the number of grid cells per zone thickness drops below a certain threshold, then grid stretching can significantly increase reflection coefficient $C_{\mathrm{R}}$.  Based on the present results, it is recommended to have cell sizes of at least $\lambda/10$ when combining grid stretching and absorbing layers. 

If the absorbing layer is subdivided into a sufficient number of zones $n$, then the difference between the theory solutions for different $n$  can be estimated by a Richardson-type extrapolation. Detailed information on Richardson extrapolation can be found e.g. in \cite{REFrichardson1,REFrichardson2,REFferzigerperic}.
Say
\begin{equation}
A = A_{\mathrm{h}} + \varepsilon_{\mathrm{h}} \quad ,
\label{EQAh}
\end{equation}
where $A$ is the analytical solution for $n=\infty$, $A_{\mathrm{h}}$ is the analytical solution for $n=x_{\mathrm{d}}/h$, and $\varepsilon_{\mathrm{h}}$ is the error. Let all zones have the same thickness $x_{\mathrm{d},j}=h$, with a total absorbing layer thickness of $x_{\mathrm{d}}=\sum_{j=1}^{n}x_{\mathrm{d},j}$. 
Thus when using zones of twice the thickness, i.e. $x_{\mathrm{d},j}=2h$, one obtains
\begin{equation}
A = A_{\mathrm{2h}} + \varepsilon_{\mathrm{2h}} \quad ,
\label{EQA2h}
\end{equation}
and similar for further refinement or coarsening.

Taylor-series analysis of truncation errors suggests that the error $\varepsilon_h$ is proportional to some power $p$ of the zone thickness $h$, i.e. $\varepsilon_h \propto h^p$. It follows that the error with  a twice coarser spacing is
\begin{equation}
\varepsilon_{\mathrm{2h}} = 2^{p} \varepsilon_{\mathrm{h}}\quad ,
\label{EQehconvergence}
\end{equation}
where $p$ is the order of convergence. Setting Eqs. (\ref{EQAh}) and (\ref{EQA2h}) as equal and inserting Eq. (\ref{EQehconvergence}) leads to 
\begin{equation}
\varepsilon_{\mathrm{h}} = \frac{A_{\mathrm{h}}-A_{\mathrm{2h}}}{2^{p}-1} \quad .
\label{EQeh}
\end{equation}

Insert Eq. (\ref{EQehconvergence}) into Eq. (\ref{EQeh}) written for $\varepsilon_{\mathrm{2h}}$ finally gives
\begin{equation}
p = \frac{\log \left( \frac{A_{\mathrm{2h}}-A_{\mathrm{4h}}}{A_{\mathrm{h}}-A_{\mathrm{2h}}} \right)}{\log \left( 2 \right) } \quad .
\label{EQp}
\end{equation}

Figures \ref{FIGCr0gdisc} and \ref{FIGCr0gdiscLOG} show that the deviation of $C_{\mathrm{R}}$ can differ depending on $\gamma$. Thus to estimate $\varepsilon_{\mathrm{h}}$ and $p$ in Eqs. (\ref{EQeh}) and (\ref{EQp}), set
\begin{equation}
A_{\mathrm{2h}} - A_{\mathrm{4h}} \approx \max \{ C_{\mathrm{R,2h}}(\gamma) - C_{\mathrm{R,4h}}(\gamma) \} \quad ,
\label{EQA2hA4h}
\end{equation}
\begin{equation}
A_{\mathrm{h}} - A_{\mathrm{2h}} \approx \max \{ C_{\mathrm{R,h}}(\gamma) - C_{\mathrm{R,2h}}(\gamma) \} \quad ,
\label{EQAhAh}
\end{equation}
where $C_{\mathrm{R,h}}(\gamma)$ is the reflection coefficient for zone thickness $x_{\mathrm{d},j}=h$ and forcing strength $\gamma$, and  the $\max \{X(\gamma)\}$-function delivers the maximum value of $X$ of all values $\gamma$ in the considered range $10\, \mathrm{s^{-1}} \leq \gamma \leq 10^{6}\, \mathrm{s^{-1}}$.

\begin{figure}[H]
\begin{center}
\includegraphics[width=\halffactor\linewidth]{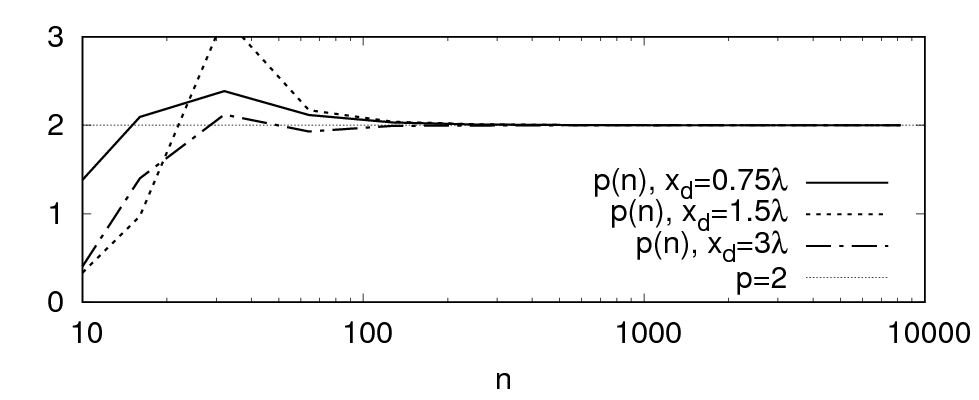} \includegraphics[width=\halffactor\linewidth]{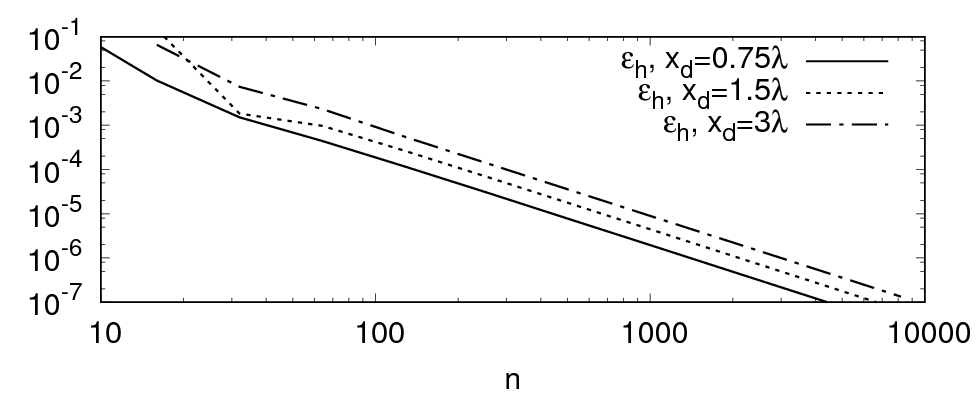} 

\end{center}
\caption{Order of convergence $p$ (left) and error estimate $\varepsilon_{\mathrm{h}}$ (right) over number of zones $n$, for the absorbing layer in Figs. \ref{FIGCr0gdisc} and \ref{FIGCr0gdiscLOG}} \label{FIGcompareDisc}
\end{figure}

Figure \ref{FIGcompareDisc} shows, exemplarily for $b(x)$ according to Eq. (\ref{EQblendEXP}),  that the analytical solution converges with $2^{\mathrm{nd}}$ order to the analytical solution for $n=\infty$.  For the blending functions investigated in Sect. \ref{SECDiscusstheoryBlend}, all curves showed $2^{\mathrm{nd}}$-order convergence (i.e. $p \rightarrow 2$ if $n \rightarrow \infty$), except constant blending ($b(x)=1$), for which the solution naturally must be exact independent of the number of zones. 
 It is out of the scope of this work to rigorously prove that for all possible $b(x)$ the order of convergence will be at least $p=2$, so this is left for future research; however, the present results suggest that this is the case. 
 Figure \ref{FIGcompareDisc} also  shows that the error estimate $\varepsilon_{\mathrm{h}}$ for different $n$ decays accordingly. Thus for practical grids in flow simulations, as well as for the theory results plotted in this work, the absorbing layer performance can be assumed independent of the number $n$ of zones or grid cells.

Given the simulation setup in Sect. \ref{SECsimsetup} it is expected that, when the grid resolution increases, the wave damping behavior in the flow simulation will converge towards the solution for the specified continuous blending function $b(x)$.
Since Sect. \ref{SECres} showed that the theory from Sect. \ref{SECtheory} predicts flow simulation results with great accuracy, it is expected that the results of the theory from Sect. \ref{SECtheory}, which is based on discontinuous piece-wise constant blending $b(x)$, will converge towards the analytical solution for any given continuous blending function  $b(x)$, if the absorbing layer is subdivided into a sufficient number of zones $n$.

\section{Application of 1D-theory to 2D- and 3D-flows}
\label{SECapply1Dto2D}
\begin{figure}[H]
\includegraphics[width=\linewidth]{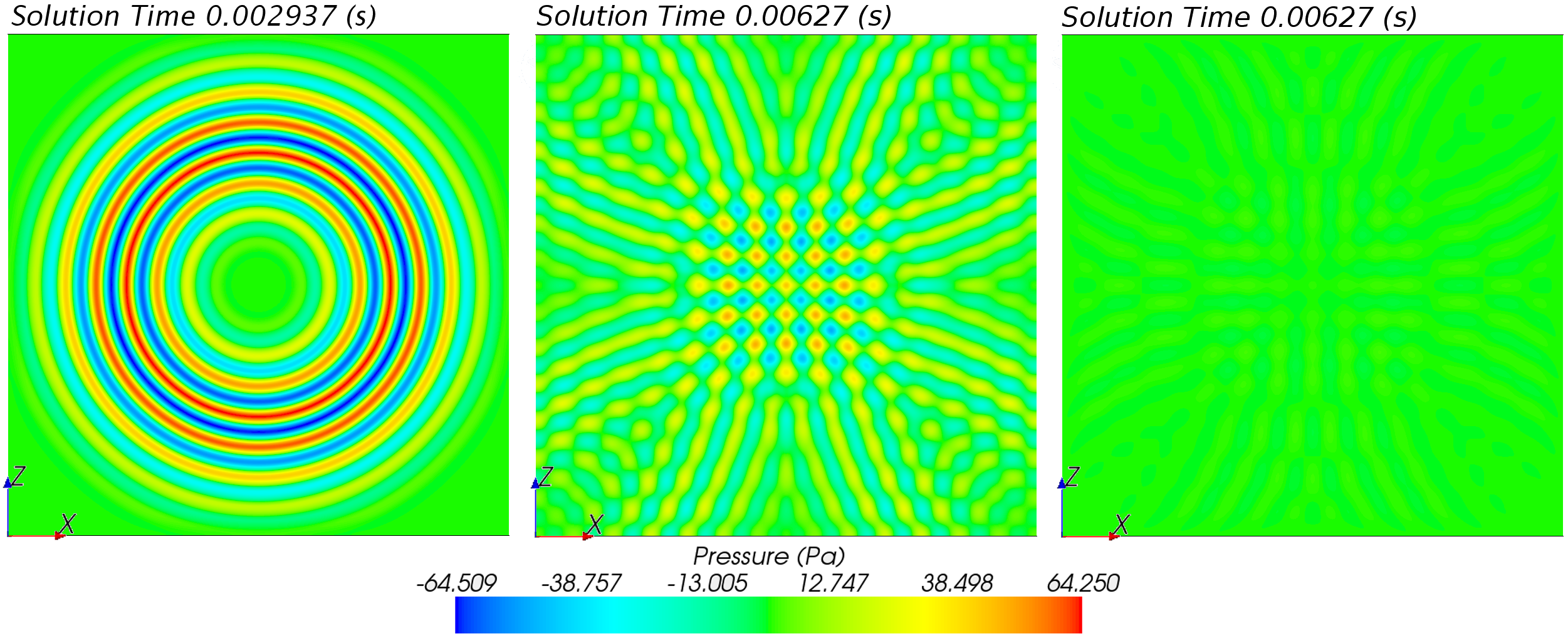}
\caption{Pressure in 2D-domain at time $t\approx 9T$ (left) and at $t = 19T$ (middle: $\gamma=5 \cdot 10^{3}\, \mathrm{s^{-1}}$ (too weak damping); right:  $\gamma=5 \cdot 10^{4}\, \mathrm{s^{-1}}$ ($\approx$ optimum damping))} 
\label{FIGB2P}
\end{figure}

The wave damping behavior in two or more dimensions can be described using 1D-theory with good approximation in many practical cases. For example, waves are often reflected at or generated by bodies within the domain, and are radiated as more or less circular waves in all directions. As a test case for such a behavior, 2D- and 3D-flow simulations are conducted using a point-like wave source. In the 2D-simulations, the source is placed in the center of a quadratic domain filled with water as shown in Fig. \ref{FIGB2P}; the domain dimensions are $0 \leq x,z \leq 16\lambda$. In the 3D-simulations, the source is placed in a center of a cubic domain with dimensions $0 \leq x,y,z \leq 16\lambda$ with all domain sides set as symmetry boundary conditions.

Waves are generated for the first $8T$ of simulation time by introducing a mass source term in Eq. (\ref{EQconti}) for $3.95\, \mathrm{m}\leq x,y,z \leq 4.05\, \mathrm{m}$ as
\begin{equation}
q_{\mathrm{c}} =
  \begin{cases}
    10\cdot \sin (-\omega t)  \cdot  \cos ^{2}(\frac{\pi}{2} + \frac{\pi}{2}\frac{t}{4T})  \, \mathrm{\frac{1}{s}}  & \quad \text{if } t \leq 4T\\
    10\cdot \sin (-\omega t)  \cdot  \cos ^{2}(\frac{\pi}{2}\frac{ t-4T}{4T}) \, \mathrm{\frac{1}{s}}   & \quad \text{if } 4T < t \leq 8T \\
    0 & \quad \text{if } t > 8T\\
  \end{cases} \quad ,
\label{EQinlet8sinewindowB2}
\end{equation}
with time $t$  and angular wave frequency $\omega$.
This produces a wave packet with period $T = 0.00033\, \mathrm{s}$ and wavelength $\lambda \approx 0.5\, \mathrm{m}$. 

To each $x$- and $z$-normal boundary, an absorbing layer according to Eq. (\ref{EQmomdamp}) is attached as
\begin{equation}
q_{x} = -\gamma \left( 1-\frac{\tilde{x}}{x_{\rm d}} \right)^{2} u \quad , \quad \quad q_{z} = -\gamma \left( 1-\frac{\tilde{z}}{x_{\rm d}} \right)^{2} w \quad .
\label{EQB2qx1}
\end{equation}

In the 3D-simulations, there is further an absorbing layer applied to each $y$-normal boundary with
\begin{equation}
q_{y} = -\gamma \left( 1-\frac{\tilde{y}}{x_{\rm d}} \right)^{2} v \quad ,
\label{EQB2qx3}
\end{equation}
with forcing strength $\gamma$, cartesian velocity components $u$, $v$, and $w$, layer thickness $x_{\mathrm{d}}=1\lambda$ and quadratic blending, where $\tilde{x}$, $\tilde{y}$, and $\tilde{z}$ give the shortest distance to the nearest $x$-, $y$-, and $z$-normal boundary. Simulations are run for different values of $\gamma$. The reflection coefficient is calculated using Eq. (\ref{EQCrB2}).

The time step is $\Delta t = 1.65\cdot 10^{-6}\, \mathrm{s} = T/200$ and $6$ iterations are performed per time step. The simulated time interval is $0\leq t \leq 0.00627 \, \mathrm{s} = 19T$. The wave is reflected either once, when incidence angle measured from the boundary normal is $ \lesssim 27\, \mathrm{deg} $, or twice for larger angles.
In the 2D-simulations, a cell size of $\Delta x = 0.00625\, \mathrm{m} \approx \lambda/80$ in both $x$- and $z$-direction is used, which results in a total of $ \approx 1.6\cdot 10^{6} $ cells. In the 3D-simulations, the grid was chosen substantially coarser to reduce the computational effort. Polyhedral cells with diameter of  $\approx 0.02\, \mathrm{m} \approx \lambda/25$ were used, resulting in $\approx 71.2\cdot 10^{6}$ cells; although the resulting 3D-grid is rather coarse, it was considered sufficient for the present purposes, i.e. to demonstrate that 1D-theory has its value also for setting up absorbing layer parameters in 2D- and 3D-simulations. The time to run the parallel simulations was in the order of several hours (2D) or days (3D) with $12$  $2.6\, \mathrm{GHz}$ processors.

\begin{figure}[H]
\includegraphics[width=\halffactor\linewidth]{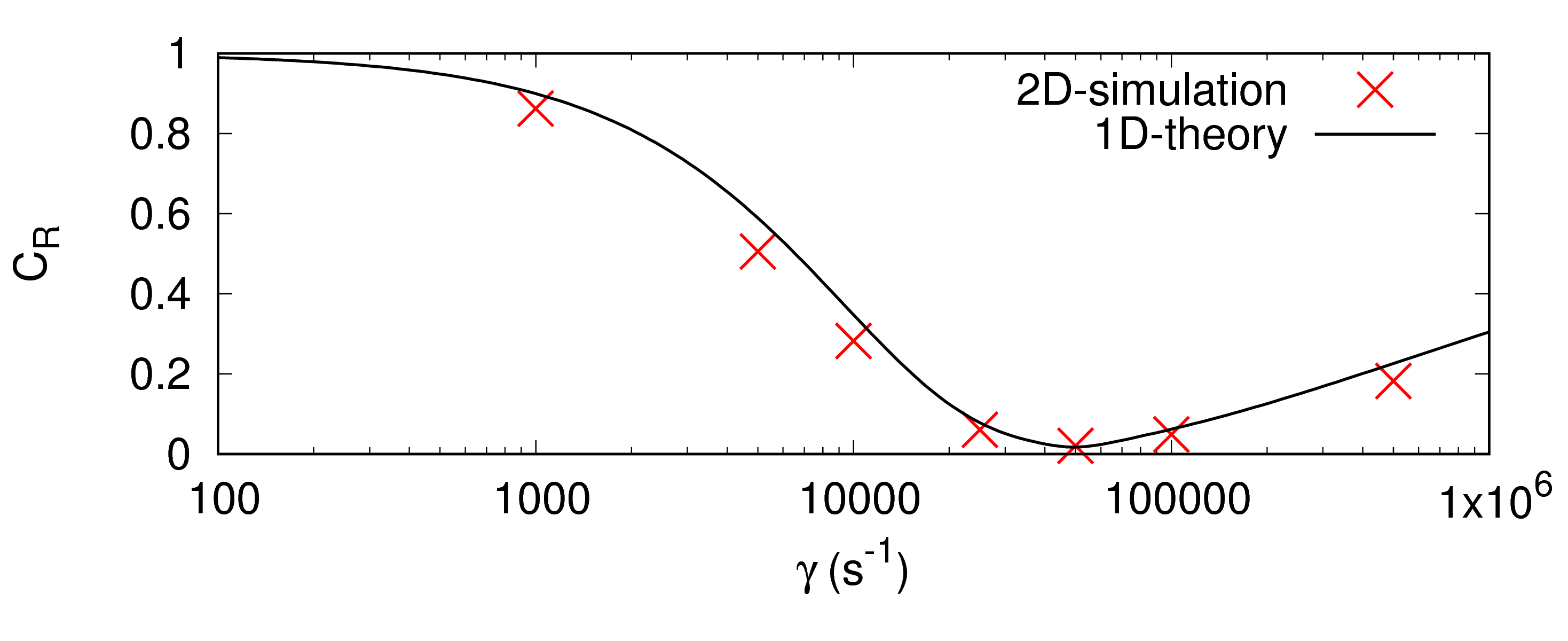}
\includegraphics[width=\halffactor\linewidth]{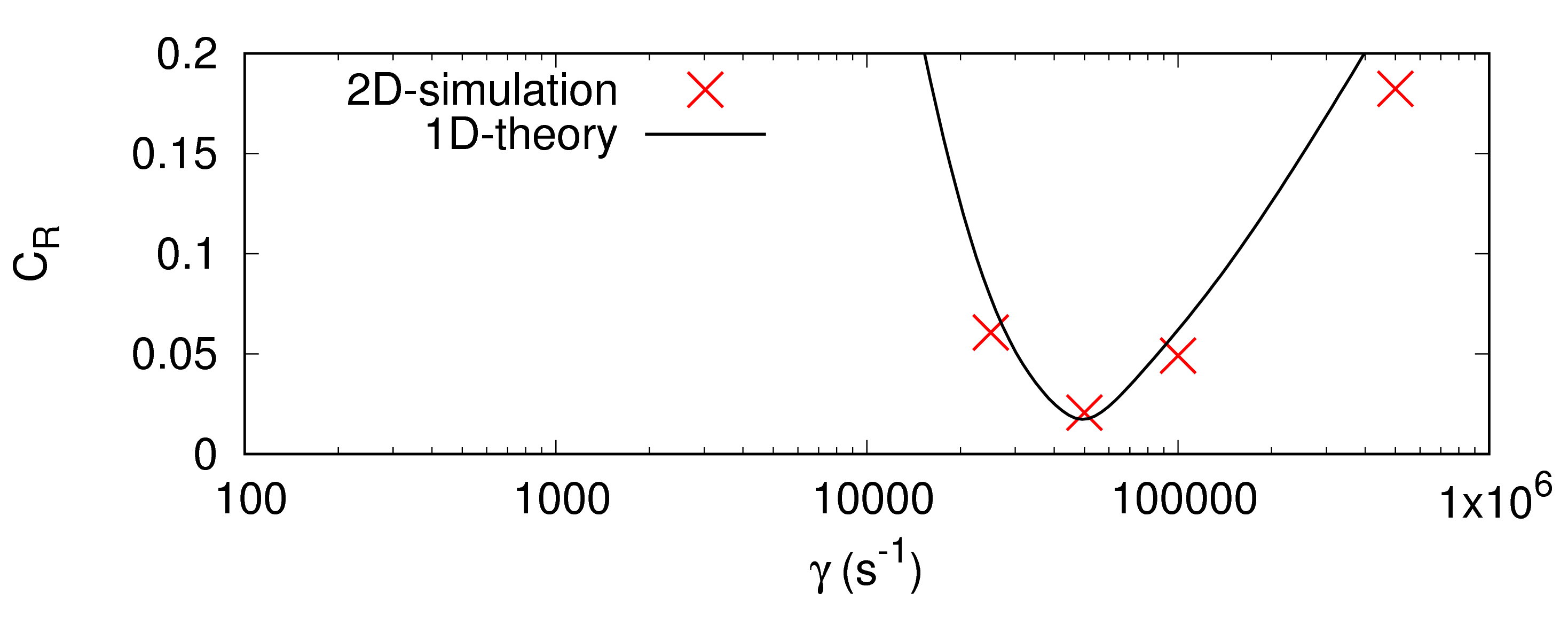}
\caption{Reflection coefficient $C_{\mathrm{R}}$ over forcing strength $\gamma$ for 2D-simulation and 1D-theory from Sect. \ref{SECtheory}} 
\label{FIGB2CR}
\end{figure}

\begin{figure}[H]
\includegraphics[width=\halffactor\linewidth]{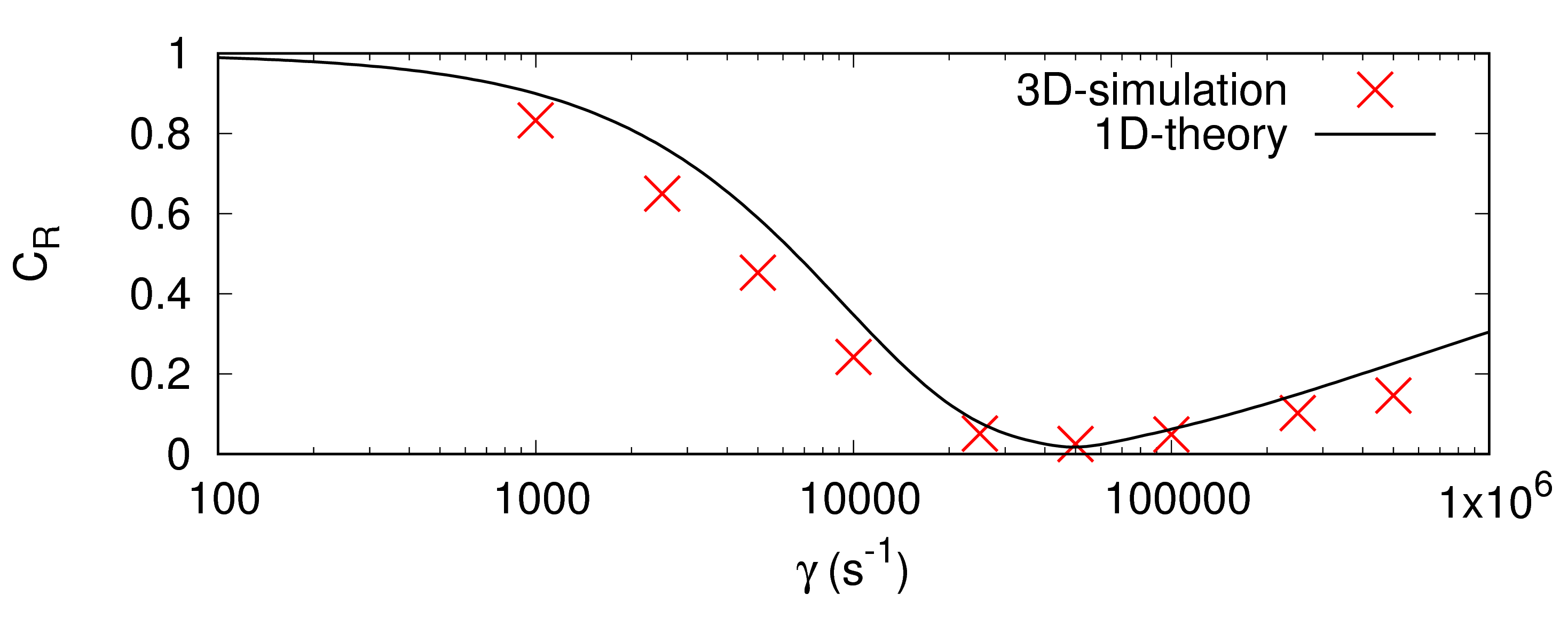}
\includegraphics[width=\halffactor\linewidth]{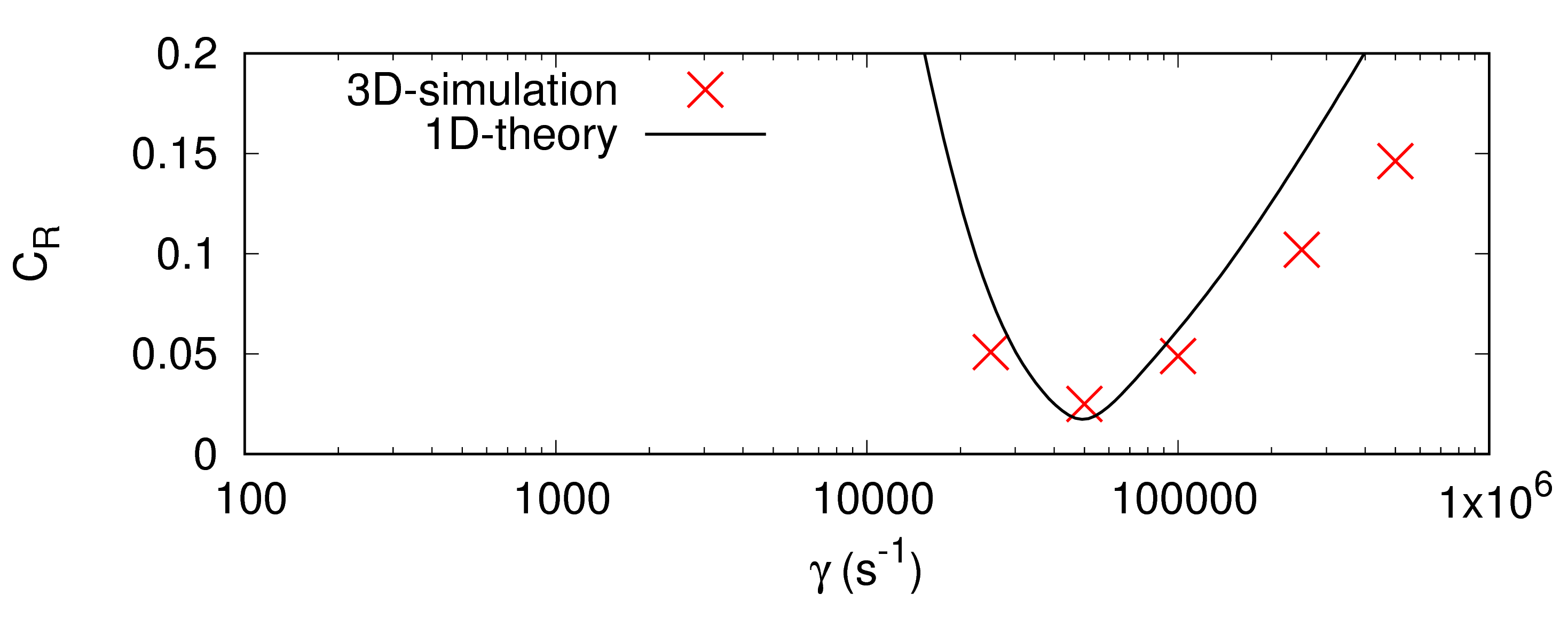}
\caption{Reflection coefficient $C_{\mathrm{R}}$ over forcing strength $\gamma$ for 3D-simulation and 1D-theory from Sect. \ref{SECtheory}} 
\label{FIGC1CR}
\end{figure}

Figures \ref{FIGB2CR} and \ref{FIGC1CR} show that the optimum choice for $\gamma $ and the resulting reflection coefficients correspond well with predictions by the 1D-theory from Sect. \ref{SECtheory}. This indicates that the damping behavior does not change substantially for moderate deviations of the wave incidence angle relative to the boundary normal (say $ \lesssim 30\, \mathrm{deg} $). Further, for larger incidence angles the waves are reflected twice before propagating back into the domain; this is expected to improve the damping since the waves encounter two absorbing layers, and at least one of the two reflections must occur with incidence angle $\leq 45\, \mathrm{deg} $; this explains why most simulation results for $C_{\mathrm{R}}$ are slightly lower than 1D-theory predictions. Thus when changing the position of the wave source, or when extending the 2D- or 3D-domains in $x$-, $y$-, or $z$-direction to achieve rectangular or cuboid domain shapes,  comparable results can be expected. 
Tuning the absorbing layer parameters according to 1D-theory predictions can therefore be considered sufficiently accurate for many practical 2D- and 3D-flow simulation problems.

However, if the wave source were positioned very close to the domain boundaries, or else if the emitted sound intensity would vary substantially with propagation direction, then parameter settings and accuracy of the 1D-theory should be assessed  using 2D- and 3D-theory as given in Sects. \ref{SECtheory2D} and \ref{SECres2D}.

\section{2D- and 3D-Theory}
\label{SECtheory2D}

In 2D- and 3D-flows the reflection coefficient may differ locally, for example due to different wave incidence angles as discussed in Sect. \ref{SECapply1Dto2D}. To accurately predict both the global reflection coefficient $C_{\mathrm{R}}$ and the flow inside the domain as for the 1D-case in Sect. \ref{SECres}, the theory would need to take into account the wave evolution in the whole domain, including the influence of reflecting objects placed within the domain.
This makes an accurate theoretical description for every possible 2D-case difficult.

\begin{figure}[H]
\includegraphics[width=0.98\linewidth]{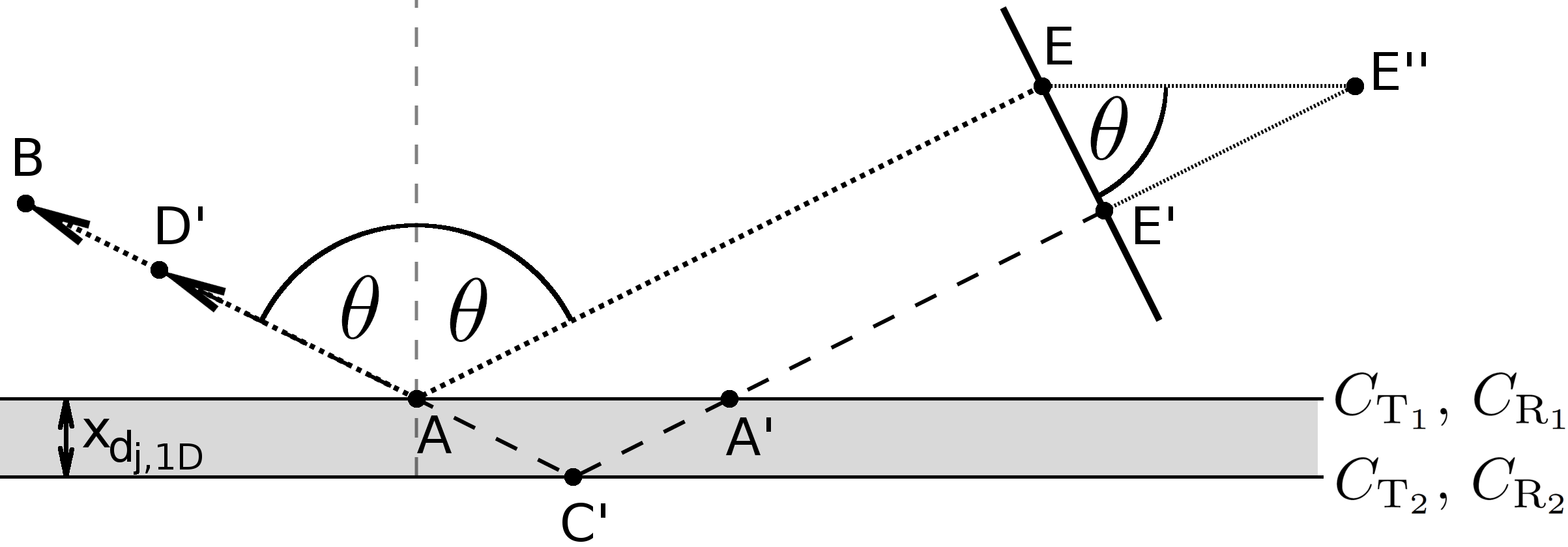}
\caption{Wave crest ( $\overline{EE'}$) propagates with incidence angle $\theta$ towards absorbing layer (shaded gray, only the first zone is depicted); given are the paths for  reflection at the layer entrance for the crest segment close to $E$  (path $\overline{EAB}$) and for reflection at the interface to the second zone for the crest segment close to $E'$ (path $\overline{E'C'D'}$); both reflected paths lie on top of each other and create interference;  distances $\overline{AC'}=x_{\mathrm{d}_{j,\mathrm{2D}}}$ and  $\overline{AA'} = \overline{EE''} = 2x_{\mathrm{d}_{j,\mathrm{2D}}}\sin \theta$ give the distance shift between reflected waves at two adjacent layer interfaces as  $\overline{E'E''} = 2x_{\mathrm{d}_{j,\mathrm{2D}}}\sin^{2} \theta$  } 
\label{FIGtheory2D2D}
\end{figure}

Therefore this section follows a more practical approach:
The 1D-theory is modified to describe 2D and 3D waves that enter the absorbing layer at an angle $\theta$ as shown in Fig. \ref{FIGtheory2D2D}. Thus the influence of  $\theta$ on $C_{\mathrm{R}}$ is obtained, and it is possible to judge whether 1D-theory predictions can be used with confidence for the 2D- and 3D-case as Sect. \ref{SECapply1Dto2D} suggests, and roughly to what extent the 1D-theory predictions will be conservative or not. 

For the theory presented in this section, the following modifications to  1D-theory are made. The incoming waves are approximated as plane waves, assuming that the waves are generated sufficiently far from the absorbing layer and that a comparatively thin segment of the absorbing layer is considered.
The waves enter the absorbing layer at an incidence angle $\theta$ between wave propagation direction and the layer-normal vector as shown in Fig. \ref{FIGtheory2D2D}; by looking at the plane spanned by these two vectors, every 3D-problem is reduced to a 2D problem. Hence in the following, the subscript '$ _{2D} $' denotes the modified theory which is applicable to 2D- and 3D-flows.
As in the 1D-case in Sect. \ref{SECtheory}, the absorbing layer is composed of several zones with constant damping, and the shortest distance between two neighboring zone interfaces $j$ and $j+1$ is $ x_{\mathrm{d}_{j\mathrm{,1D}}}$.
Waves can be partially reflected at each interface between two zones, and interference occurs between those reflected wave components which travel along the same path.

Two mechanisms modify the reflection behavior in contrast to the 1D-case. First, the wave travels a larger distance through the absorbing layer, which acts like an increase of  the layer thickness $x_{d}$ proportional to the increase in propagation distance within the layer; thus each zone acts on the wave as if it had thickness
\begin{equation}
x_{\mathrm{d}_{j\mathrm{,2D}}} = \frac{x_{\mathrm{d}_{j\mathrm{,1D}}}}{\cos (\theta)} \quad ,
\label{EQxdi2d}
\end{equation}
with  thickness  $x_{\mathrm{d}_{j\mathrm{,1D}}} $ of zone $j$ and incidence angle $ |\theta| \leq 90\, \mathrm{deg}$.

Second, the amount of destructive interference depends on the phase shift between the reflected waves. 
In 1D-theory the distance shift between the crests of waves reflected at two adjacent zone interfaces $j$ and $j+1$ is always $2 x_{\mathrm{d}_{j\mathrm{,1D}}}$;
for the 2D-case, the distance shift is $\leq 2 x_{\mathrm{d}_{j\mathrm{,2D}}}$, since a wave reflected at a certain location will interfere with wave reflections which occur at neighboring positions as illustrated in Fig. \ref{FIGtheory2D2D}.

Trigonometric considerations (see Fig. \ref{FIGtheory2D2D}) show that the distance shift is
\begin{equation}
\Delta x = 2 x_{\mathrm{d}_{j\mathrm{,2D}}}   \sin ^{2} \theta  \quad .
\label{EQdx2d}
\end{equation}

For reflections at the interface between zones $j$ and $j+1$, this leads to a phase shift for the reflected wave component in zone $j+1$  by factor
\begin{equation}
E = \mathrm{e}^{ i k_{0} \left( -\Delta x \right)} \quad ,
\label{EQEj2d}
\end{equation}
where $k_{0}$ is the wave number outside the absorbing layer.

Therefore, the 2D-problem can be reduced to a 1D-problem to save computational effort: The reflection coefficients for the 2D-case in Fig. \ref{FIGtheory2D2D} are equivalent to those obtained using 1D-theory, if the thickness of the zones is set to $x_{\mathrm{d}_{j\mathrm{,2D}}}$ instead of $x_{\mathrm{d}_{j\mathrm{,1D}}}$, and additionally the phase shift between waves reflected at adjacent zones is adjusted according to Eq. (\ref{EQEj2d}). 

Consider two adjacent zones $j$ and $j+1$. The particle displacements for the equivalent 1D-case are

\begin{align}
 \chi_{j} =  \chi_{0}  \left( \prod_{n=0}^{j-1} C_{\mathrm{T}_{n}} \right) \biggl[ \mathrm{e}^{i \left( \sum_{n=1}^{j-1} k_{n} x_{\mathrm{d}_{n,\mathrm{2D}}} + k_{j} \left( x -  \sum_{n=1}^{j-1} x_{\mathrm{d}_{n,\mathrm{2D}}  }   \right) \right)} \nonumber \\ 
  -  C_{\mathrm{R}_{j}} \mathrm{e}^{i \left(  \sum_{n=1}^{j-1} k_{n} x_{\mathrm{d}_{n,\mathrm{2D}}}  + k_{j}2x_{\mathrm{d}_{j,\mathrm{2D}}} - k_{j} \left(x -  \sum_{n=1}^{j-1} x_{\mathrm{d}_{n,\mathrm{2D}} }   \right) \right)} 
\biggr]  \quad ,
\label{EQXj2D}
\end{align} 
\begin{align}
 \chi_{j+1} =  \chi_{0}  \left( \prod_{n=0}^{j} C_{\mathrm{T}_{n}} \right) \biggl[ \mathrm{e}^{i \left( \sum_{n=1}^{j} k_{n} x_{\mathrm{d}_{n,\mathrm{2D}}} + k_{j+1} \left( x -  \sum_{n=1}^{j} x_{\mathrm{d}_{n,\mathrm{2D}}  }   \right) \right)} \nonumber \\ 
  - E \cdot  C_{\mathrm{R}_{j+1}} \mathrm{e}^{i \left(  \sum_{n=1}^{j} k_{n} x_{\mathrm{d}_{n,\mathrm{2D}}}  + k_{j+1}2x_{\mathrm{d}_{j+1,\mathrm{2D}}} - k_{j+1} \left(x -  \sum_{n=1}^{j} x_{\mathrm{d}_{n,\mathrm{2D}} }   \right) \right)} 
\biggr]  \quad .
\label{EQXjp12D}
\end{align} 

Requiring that at the interface between zones $j$ and $j+1$ should hold $ \left[ \chi_{j} =  \chi_{j+1} \right]_{x=\sum_{n=1}^{j} x_{\mathrm{d}_{n}}}$ and  $ \left[ \chi_{x, j} =  \chi_{x, j+1} \right]_{x=\sum_{n=1}^{j} x_{\mathrm{d}_{n}}}$,  transmission and reflection coefficients for interface $j$ can be derived as
\begin{framed}
\begin{align}
 C_{\mathrm{T}_{j}} =  \frac{1  -  C_{\mathrm{R}_{j}}}{ 1 - E \cdot C_{\mathrm{R}_{j+1}}  \mathrm{e}^{i \left(  k_{j+1}2x_{\mathrm{d}_{j+1,\mathrm{2D}}}  \right) }} 
 \quad ,
\label{EQCt}
\end{align} 

\begin{align}
  C_{\mathrm{R}_{j}} = \frac{ k_{j+1}\beta_{j+1} -  k_{j}   }{  k_{j+1}\beta_{j+1} +  k_{j}   }
 \quad ,
\label{EQCRj}
\end{align} 
with
\begin{align}
\beta_{j+1} = \frac{  1 +  E \cdot C_{\mathrm{R}_{j+1}}  \mathrm{e}^{i  \left( k_{j+1} 2 x_{\mathrm{d}_{j+1,\mathrm{2D}}} \right) }  }{  1 -  E \cdot C_{\mathrm{R}_{j+1}}  \mathrm{e}^{i  \left( k_{j+1} 2 x_{\mathrm{d}_{j+1,\mathrm{2D}}} \right) } }  \quad .
\label{EQbetajp1}
\end{align} 
\end{framed}
If the absorbing layer starts at zone $1$, the 'global' reflection coefficient $  C_{\mathrm{R}} $ for the 2D-case is 
\begin{framed}
 \begin{align}
   C_{\mathrm{R}} = | C_{\mathrm{R}_{1}} | = \sqrt{\mathrm{Re}\{C_{\mathrm{R}_{1}}\}^{2} + \mathrm{Im}\{C_{\mathrm{R}_{1}}\}^{2}}
 \quad ,
\label{EQCRglobal}
\end{align} 
\end{framed}
where $ \mathrm{Re}\{ X\} $ and $ \mathrm{Im}\{ X\} $ denote the real and the imaginary part of the complex number $X$.

\section{2D-Results}
\label{SECres2D}

To verify the 2D-theory from Sect. \ref{SECtheory2D}, 2D-flow simulations are performed based on the setup in Sect. \ref{SECapply1Dto2D}, except that domain dimensions are $   -43\lambda \approx -21.5\, \mathrm{m} \leq x \leq 0\, \mathrm{m}$ and $ 0\, \mathrm{m} \leq z \leq 12.9\, \mathrm{m} \approx 26\lambda$ as shown in Fig. \ref{FIGB3P}. 
The mass source term from Eq. (\ref{EQinlet8sinewindowB2}) is applied for $-0.05\, \mathrm{m}\leq x \leq 0\, \mathrm{m}$ and $12.85\, \mathrm{m}\leq z \leq 12.9\, \mathrm{m}$ to generate circular waves, and there is a symmetry boundary condition at all walls to save cells. 
This produces a wave packet with period $T = 0.00033\, \mathrm{s}$ and wavelength $\lambda \approx 0.5\, \mathrm{m}$.
The waves are reflected at most one time at boundary $z=0\, \mathrm{m}$, to which an absorbing layer is attached according to Eq. (\ref{EQmomdamp})
\begin{equation}
q_{x} = -\gamma b(z) u \quad , \quad q_{z} = -\gamma b(z) w \quad ,
\label{EQB3qx}
\end{equation}
with velocities $u$ and $w$, forcing strength $\gamma$ and quadratic blending $b(z)=( (x_{\rm d} - z) / x_{\rm d} )^{2}$. Thus waves are reflected for continuous incidence angles in the range $0\leq \theta \lesssim 50\, \mathrm{deg}$.
Time step and cell sizes are as in Sect. \ref{SECapply1Dto2D}. The simulated time interval is $0\leq t \leq 0.016995 \, \mathrm{s} = 51.5T$. 
For volume $V_{\theta}$ of a thin domain slice along the paths of waves reflected at $\theta = 0\, \mathrm{deg}$, $11.25\, \mathrm{deg}$, $22.5\, \mathrm{deg}$, $33.75\, \mathrm{deg}$ and $45\, \mathrm{deg}$ as indicated in Fig. \ref{FIGB3P}, the energy of the reflected waves is integrated.
 Relating these energies for the damped ($\gamma \neq 0$) and undamped ($\gamma = 0$) case, and evaluating Eq. (\ref{EQCrB2}) at $t=51.5T$ gives the reflection coefficient $C_{\mathrm{R}}$ for each angle $\theta$.

\begin{figure}[H]
\includegraphics[width=\halffactor\linewidth]{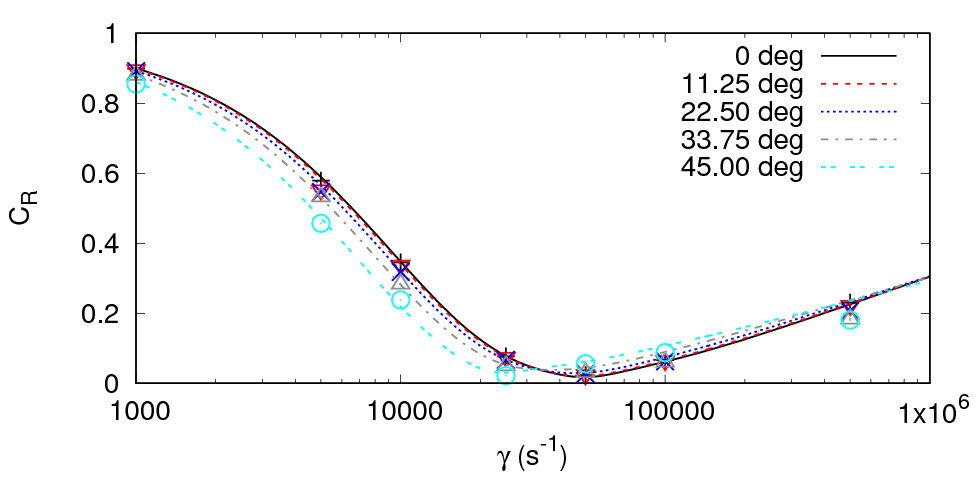}
\includegraphics[width=\halffactor\linewidth]{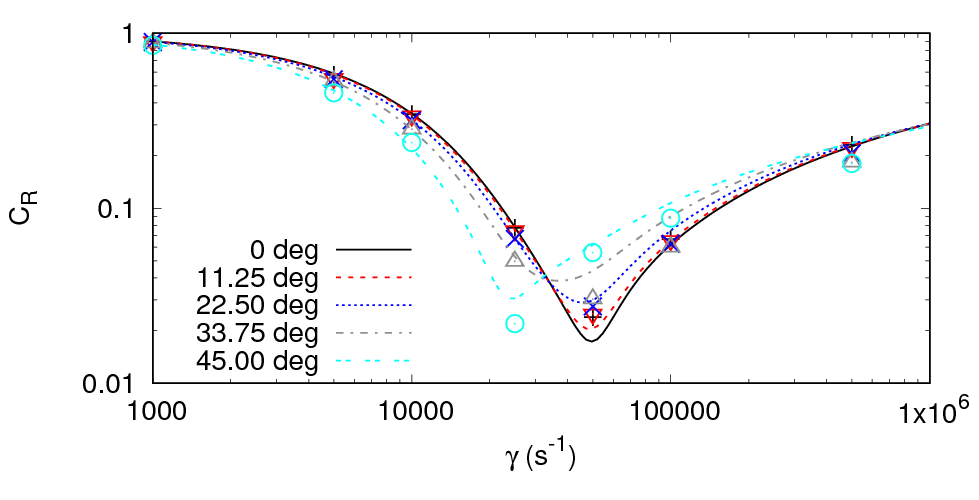}
\caption{ Reflection coefficient over forcing strength $\gamma$ for different incidence angles $\theta$; for simulation at $ 0\, \mathrm{deg} $ ($ + $), $ 11.25\, \mathrm{deg} $ ($ \bigtriangledown $), $ 22.5\, \mathrm{deg} $ ($ \times $), $ 33.75\, \mathrm{deg} $ ($\bigtriangleup  $), $ 45\, \mathrm{deg} $ ($ \bigcirc $), and for 2D-theory (dashed lines)} 
\label{FIGB3P2}
\end{figure}

\begin{figure}[H]
\includegraphics[width=\halffactor\linewidth]{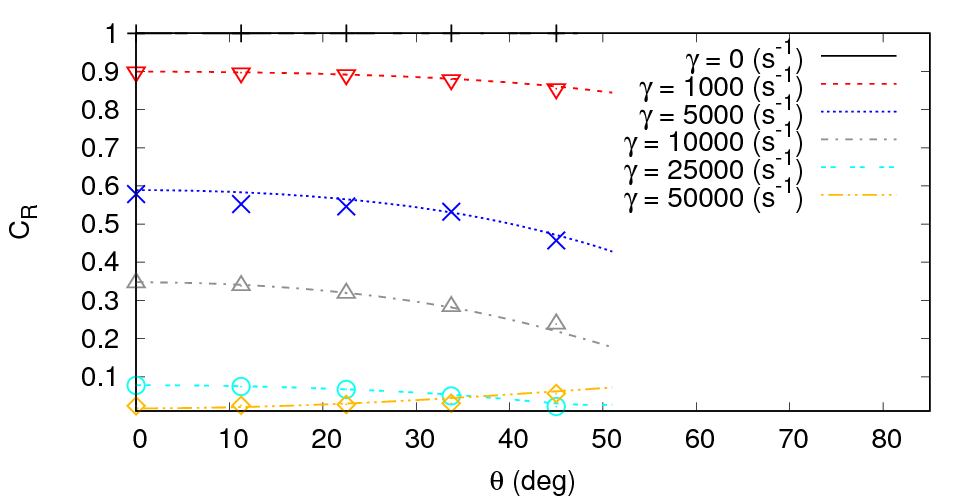}
\includegraphics[width=\halffactor\linewidth]{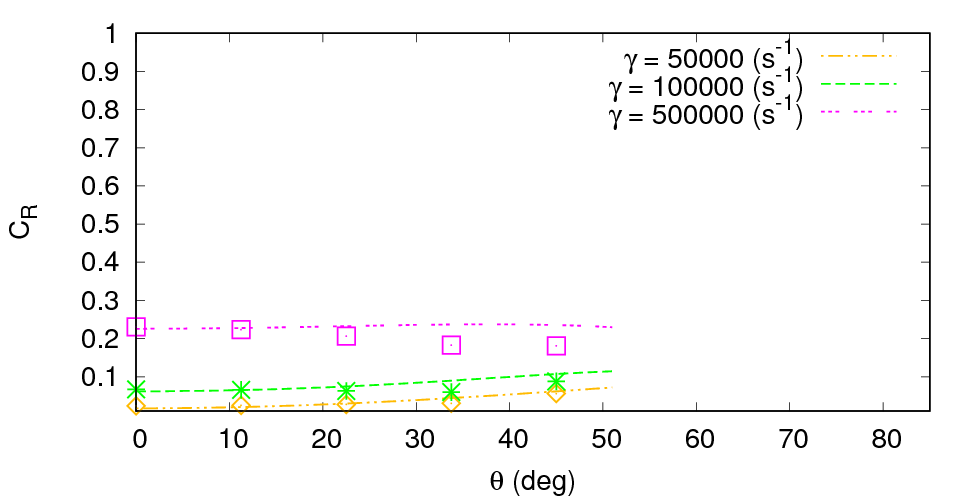}\\
\includegraphics[width=\halffactor\linewidth]{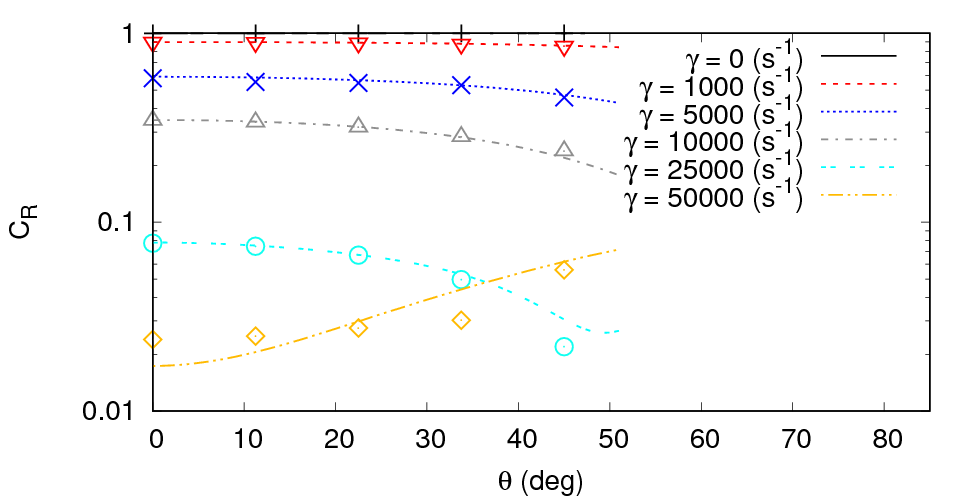}
\includegraphics[width=\halffactor\linewidth]{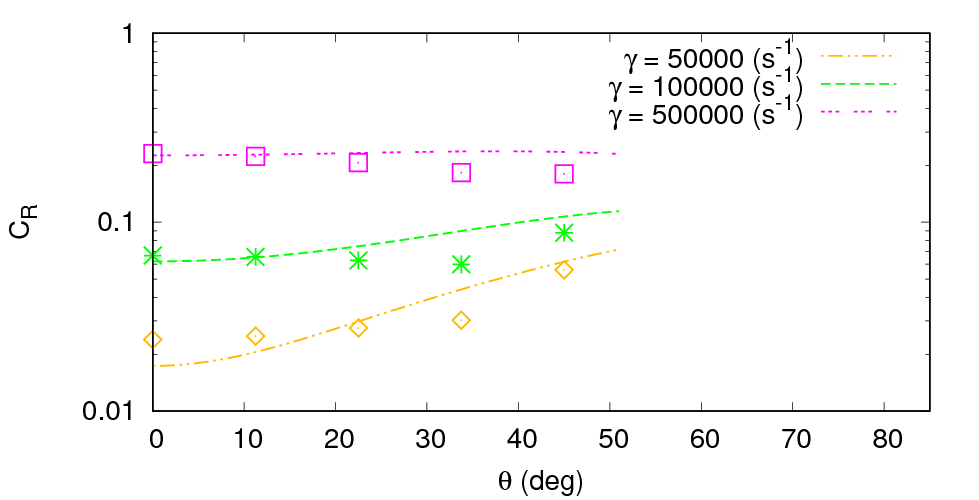}
\caption{Reflection coefficient $C_{\mathrm{R}}$ over incidence angle $\theta$ for different $\gamma$;   for $\gamma \lesssim \gamma_{\mathrm{opt}}$ (left) and for $ \gamma \gtrsim \gamma_{\mathrm{opt}}  $ (right); for simulation with $ \gamma = 0\, \mathrm{\frac{1}{\mathrm{s}}} $ ($ + $), $ \gamma = 1000\, \mathrm{\frac{1}{\mathrm{s}}} $ ($ \bigtriangledown $), $ \gamma = 5000\, \mathrm{\frac{1}{\mathrm{s}}} $ ($ \times $), $ \gamma = 10 000\, \mathrm{\frac{1}{\mathrm{s}}} $ ($\bigtriangleup  $), $ \gamma =25 000\, \mathrm{\frac{1}{\mathrm{s}}} $ ($ \bigcirc $), $ \gamma =50 000\, \mathrm{\frac{1}{\mathrm{s}}} $ ($ \diamond $),  $ \gamma =100 000\, \mathrm{\frac{1}{\mathrm{s}}} $ (\ding{103}),  $ \gamma =500 000\, \mathrm{\frac{1}{\mathrm{s}}} $ ($ \Box $), and for 2D-theory (dashed lines)} 
\label{FIGB3CR3}
\end{figure}

\begin{figure}[H]
\includegraphics[width=0.7\linewidth]{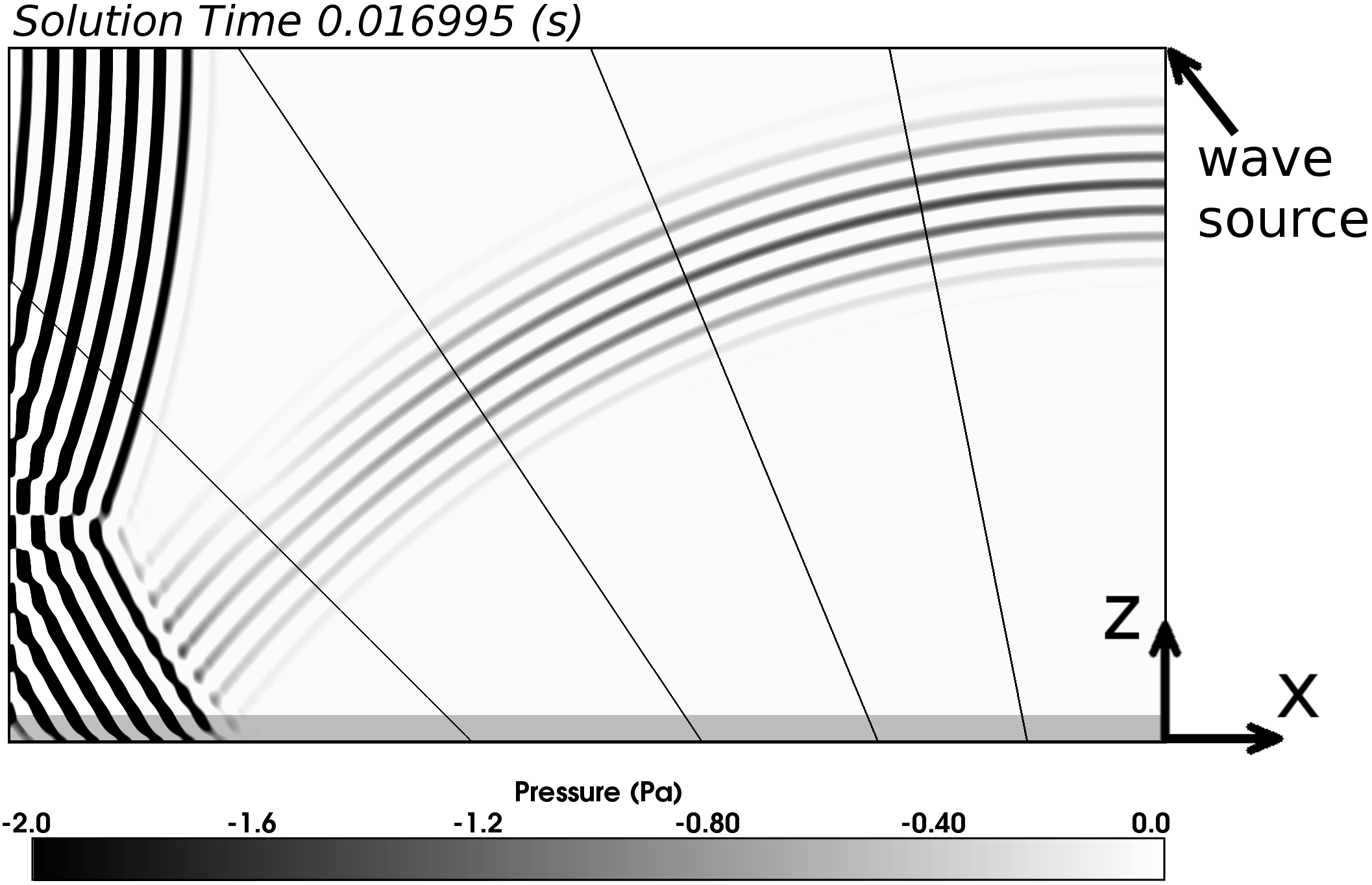}
\caption{Pressure in domain at $t=51.5T$ for $\gamma=25 000\, \mathrm{s^{-1}}$; straight lines within domain denote the paths of waves reflected at incidence angles $\theta =  11.25\, \mathrm{deg}, 22.5\, \mathrm{deg}, 33.75\, \mathrm{deg}, 45\, \mathrm{deg}$ relative to boundary normal; the absorbing layer (shaded gray, thickness $x_{\mathrm{d}}=1\lambda$) is attached to boundary $z=0$; the reflection coefficient decreases from $0\, \mathrm{deg}$ until $\approx 45\, \mathrm{deg}$, and increases for larger incidence angles, as predicted by 2D-theory (see Fig. \ref{FIGB3CR3})} 
\label{FIGB3P}
\end{figure}

First, flow simulations are run with layer thickness $x_{\mathrm{d}}=1\lambda$ for different forcing strengths $\gamma$. Figures \ref{FIGB3P2} to \ref{FIGB3CR3} show that the the influence of incidence angle $\theta$ on reflection coefficient $ C_{\mathrm{R}} $ is accurately predicted up to optimum forcing strength ($0 \leq \gamma \lesssim \gamma_{\mathrm{opt}}$), and is slightly conservative for too strong damping ($\gamma_{\mathrm{opt}} \lesssim \gamma $). 
With increasing incidence angle $\theta$ the reflection coefficient $ C_{\mathrm{R}} $ decreases or stays roughly the same for most values of $\gamma$ in the range of $0 \leq \theta \leq 45\, \mathrm{deg}$; only for $\gamma$ close to its 1D-optimum value a significant increase in $ C_{\mathrm{R}} $ occurs, since the 2D-optimum value of $\gamma$ decreases when increasing $\theta$.

Further, Fig. \ref{FIGB3P} shows that for $\gamma$ slightly below 1D optimum (here: $\gamma=25000\, \mathrm{s^{-1}}$), the reflection coefficient may first decrease ($0^{\circ} \leq \theta \lesssim 45^{\circ}$), and then increase again  ($\theta \gtrsim 45^{\circ}$); this is also predicted by 2D-theory in Fig. \ref{FIGB3CR3}.

\begin{figure}[H]
\includegraphics[width=\halffactor\linewidth]{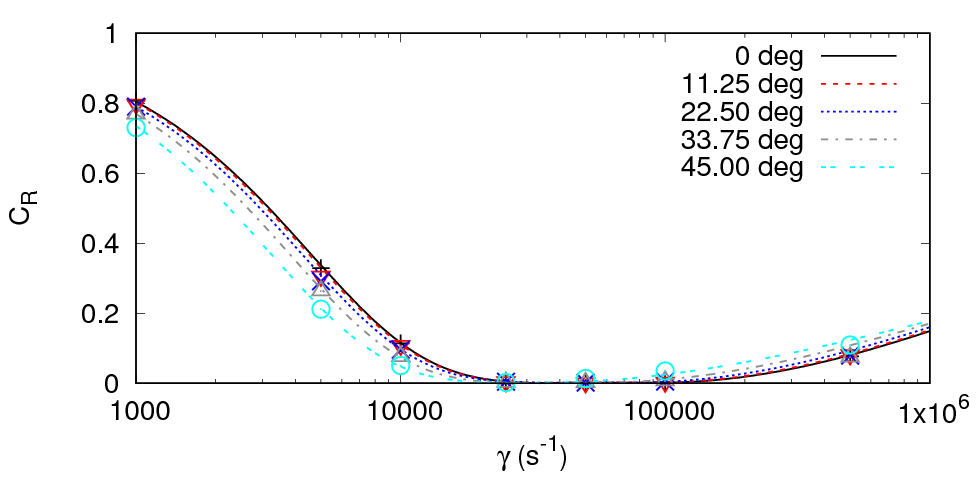}
\includegraphics[width=\halffactor\linewidth]{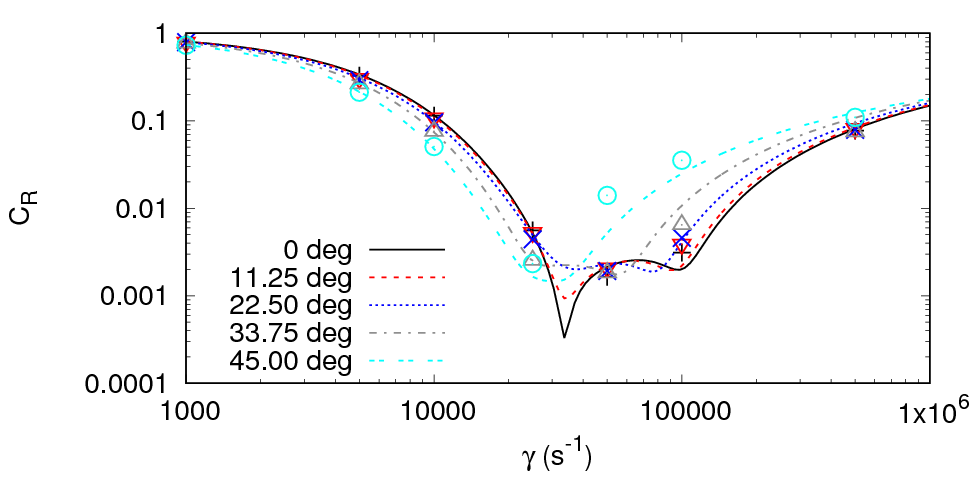}
\caption{ Reflection coefficient over forcing strength $\gamma$ for different incidence angles $\theta$; for simulation at $ 0\, \mathrm{deg} $ ($ + $), $ 11.25\, \mathrm{deg} $ ($ \bigtriangledown $), $ 22.5\, \mathrm{deg} $ ($ \times $), $ 33.75\, \mathrm{deg} $ ($\bigtriangleup  $), $ 45\, \mathrm{deg} $ ($ \bigcirc $), and for 2D-theory (dashed lines)} 
\label{FIGB5P2}
\end{figure}

\begin{figure}[H]
\includegraphics[width=\halffactor\linewidth]{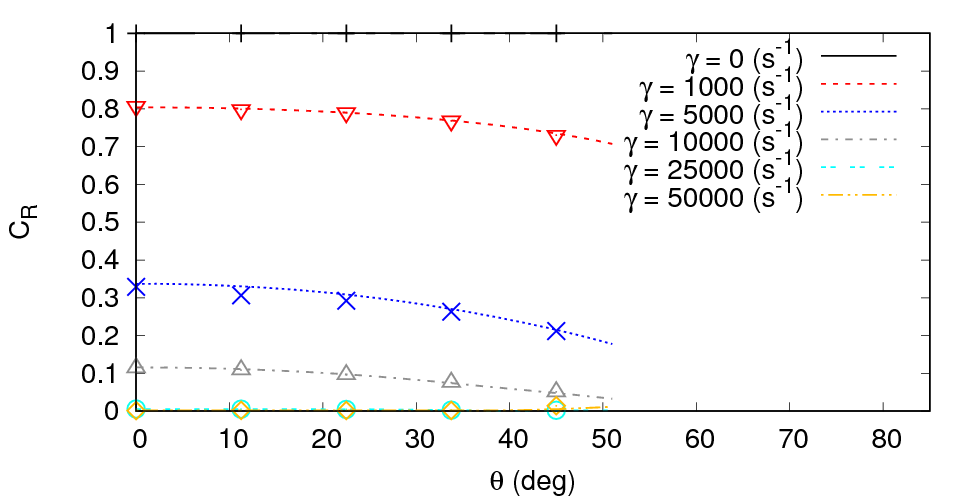}
\includegraphics[width=\halffactor\linewidth]{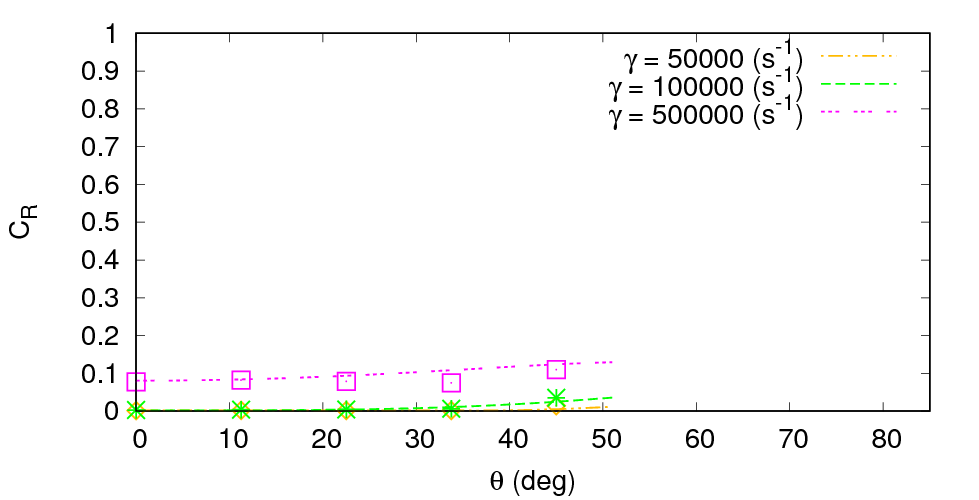}\\
\includegraphics[width=\halffactor\linewidth]{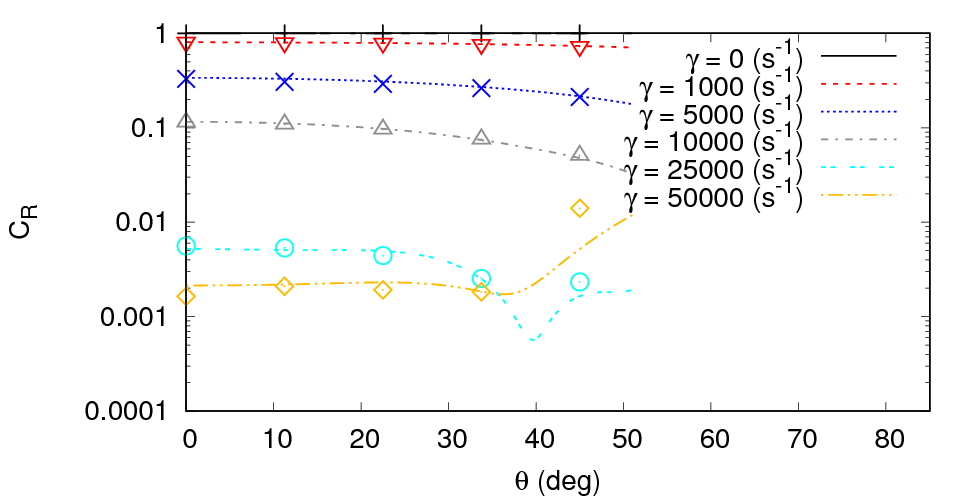}
\includegraphics[width=\halffactor\linewidth]{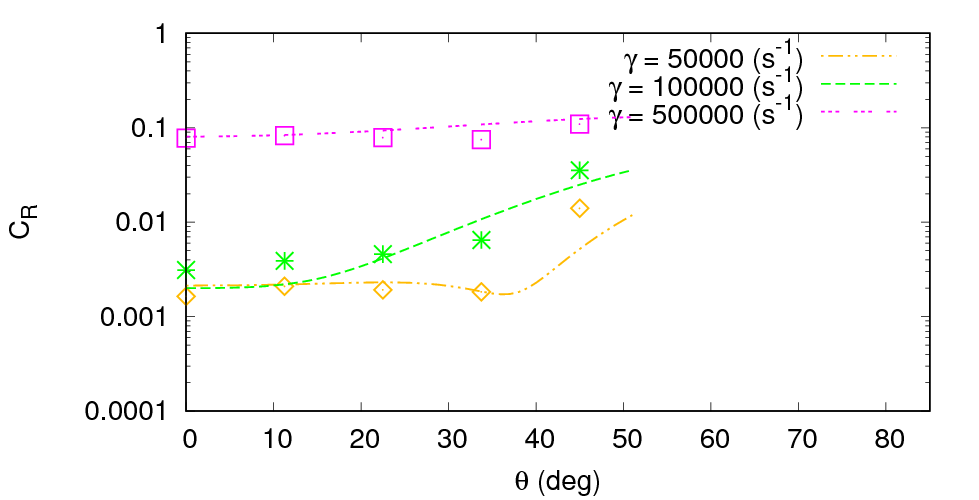}
\caption{Reflection coefficient $C_{\mathrm{R}}$ over incidence angle $\theta$ for different $\gamma$;   for $\gamma \lesssim \gamma_{\mathrm{opt}}$ (left) and for $ \gamma \gtrsim \gamma_{\mathrm{opt}}  $ (right); for simulation with $ \gamma = 0\, \mathrm{\frac{1}{\mathrm{s}}} $ ($ + $), $ \gamma = 1000\, \mathrm{\frac{1}{\mathrm{s}}} $ ($ \bigtriangledown $), $ \gamma = 5000\, \mathrm{\frac{1}{\mathrm{s}}} $ ($ \times $), $ \gamma = 10 000\, \mathrm{\frac{1}{\mathrm{s}}} $ ($\bigtriangleup  $), $ \gamma =25 000\, \mathrm{\frac{1}{\mathrm{s}}} $ ($ \bigcirc $), $ \gamma =50 000\, \mathrm{\frac{1}{\mathrm{s}}} $ ($ \diamond $),  $ \gamma =100 000\, \mathrm{\frac{1}{\mathrm{s}}} $ (\ding{103}),  $ \gamma =500 000\, \mathrm{\frac{1}{\mathrm{s}}} $ ($ \Box $), and for 2D-theory (dashed lines)} 
\label{FIGB5CR3}
\end{figure}

\begin{figure}[H]
\includegraphics[width=0.7\linewidth]{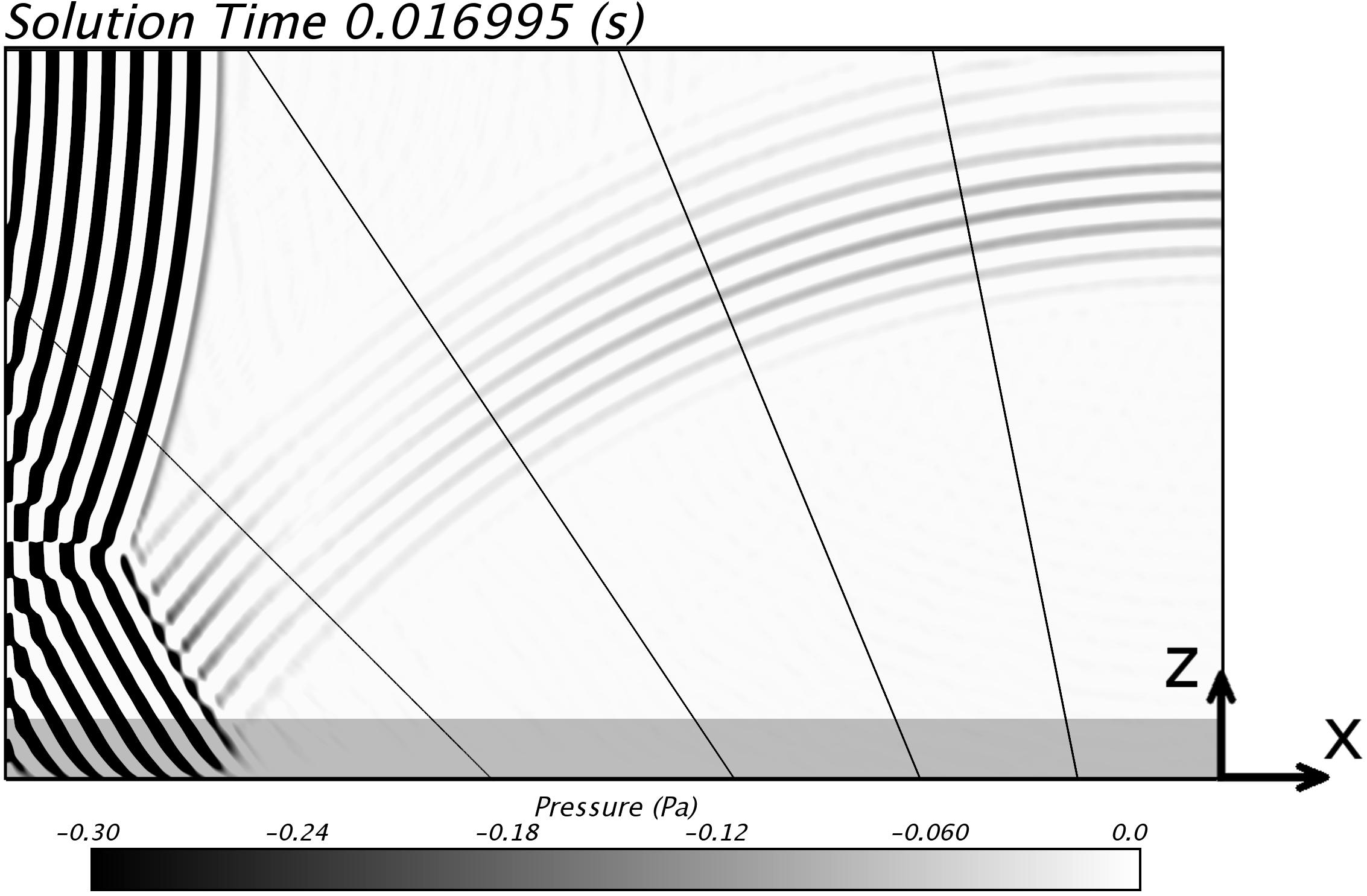}
\caption{Pressure in domain at $t=51.5T$ for $\gamma=25 000\, \mathrm{s^{-1}}$; straight lines within domain denote paths of waves reflected at incidence angles $\theta =  11.25\, \mathrm{deg}, 22.5\, \mathrm{deg}, 33.75\, \mathrm{deg}, 45\, \mathrm{deg}$, measured between wave propagation direction and boundary normal; the absorbing layer (shaded gray, thickness $x_{\mathrm{d}}=2\lambda$) is attached to boundary $z=0$; the reflection coefficient decreases from $0\, \mathrm{deg}$ until $\approx 40\, \mathrm{deg}$, and increases for larger incidence angles, as predicted by 2D-theory (see Fig. \ref{FIGB5CR3})} 
\label{FIGB5p}
\end{figure}

Second, the flow simulations are repeated with layer thickness $x_{\mathrm{d}}=2\lambda$. 
Figures \ref{FIGB5P2} to \ref{FIGB5CR3} show that, as before, the 2D-theory from Sect. \ref{SECtheory2D} satisfactorily predicts the reflection coefficients for incidence angles in the range of practical interest. Apart from improved damping due to the increased layer thickness, the trends of the curves are similar as before; again, for $\gamma$ slightly below 1D optimum (here:  $\gamma=25000\, \mathrm{s^{-1}}$) the  decrease ($0^{\circ} \leq \theta \lesssim 40^{\circ}$) and subsequent increase ($\theta\gtrsim 40^{\circ}$) of $ C_{\mathrm{R}} $ for increasing $\theta$ is captured by 2D-theory as Figs. \ref{FIGB5CR3} and \ref{FIGB5p} show.

Based on the results in this section, the following procedure is recommended for practice: When using absorbing layers that can be described in terms of the general formulation given in Sect. \ref{SECabslayer}, then 1D-theory from Sect. \ref{SECtheory} can be used to tune the absorbing layer parameters (forcing strength $\gamma$, blending function $b(x)$, layer thickness $x_{\mathrm{d}}$) to the specific problem; in most 2D- and 3D-flow problems, the prediction for optimum forcing strength and corresponding reflection coefficient will already be sufficiently accurate for practical purposes. 

If higher accuracy is needed, or if the problem is such that strong wave incidence at larger angles $\theta$ may occur, then the 2D-theory from Sect. \ref{SECtheory2D} should be used, and with the following procedure it is possible to define an upper bound for the overall reflection coefficient $ C_{\mathrm{R}} $. The results show that, for rectangular domains with absorbing layers at all boundaries, waves that are reflected twice before traveling back into the solution domain encounter an absorbing layer at least once with an angle $|\theta|\leq 45\, \mathrm{deg}$; thus they have $ C_{\mathrm{R}} $ lower or equal to 1D-theory prediction, which also holds for corners. Therefore it suffices to look at the range for $\theta$ in which reflection occurs only once, which in most cases can be determined easily; for example for the problem in Sect. \ref{SECapply1Dto2D}, where single reflection occurs for $|\theta|<30\, \mathrm{deg}$, the upper bound for the overall reflection coefficient $ C_{\mathrm{R}} $ for a given forcing strength $\gamma$ corresponds to the maximum $ C_{\mathrm{R}}(\theta) $ for 2D-theory in the range  $|\theta|<30\, \mathrm{deg}$. Since a 2D-theory prediction for one angle requires roughly the same computational effort as a 1D-theory prediction, a plot like Fig. \ref{FIGB5CR3} can be generated within a few seconds on a single core $2.6\mathrm{GHz}$ processor.
Problems in 3D can be reduced to 2D by considering the plane spanned by wave propagation direction and absorbing layer normal vector.

\section{Conclusion}

This work presented a theory to predict the reflection coefficient of absorbing layers, with main application in finite-volume based simulations of complex viscous flows as used for fundamental research in acoustics. The theory was presented for the 1D-case and extended to 2D- and 3D cases of oblique waves.

Theory predictions were compared to results from finite-volume-based flow simulations for various test cases, including regular and irregular pressure waves in liquid water and ideal gases in 1D as well as flows with oblique wave incidence in 2D and 3D. The theory predicted the optimum setup for the case-dependent parameters of the absorbing layer, as well as the corresponding reflection coefficients, with satisfactory accuracy for practical purposes. A computer code for evaluating the theory has been made available to the scientific community as free software.

The theory is based on piecewise-constant blending functions, which for a sufficient number of segments was shown to converge towards the solution for the continuous blending function. This explains why, for practical discretizations, the reduction of undesired wave reflections in the simulations was found to be independent of the order, time step size and mesh size of the chosen discretization. Writing the theory for piecewise-constant blending also has the advantage that the theory holds for any continuous or discontinuous blending function, and thus applies to many implementations in commercial and free software codes. 

The theory provides insight into the wave absorption mechanism: Wave reflections occur everywhere within the layer where the source term strength changes (i.e. when $\vec{\nabla}b(x) \neq 0$), and the absorption is largely based on destructive interference due to the phase differences of these components. 

Further, it was found that there exists no single optimum setup for the case-dependent parameters in the absorbing layer formulation. Instead, the most efficient choice of parameters will be different  depending on, for example, the intended reflection coefficient, and the frequency range and spectrum of the investigated waves. 

Topics of future research include automatic fine-tuning of the absorbing layer parameters, applications to various complex flow problems, and investigation of the prediction accuracy for damping of highly non-linear waves, including distorted and shock waves.

\section*{Acknowledgements}
The study was supported by the Deutsche Forschungsgemeinschaft (DFG).

\bibliography{}

\end{document}